\documentclass[lettersize,journal]{IEEEtran}
\usepackage{amsmath,amsfonts}
\usepackage[thmmarks,amsmath]{ntheorem}
\usepackage{algorithmic}
\usepackage{algorithm}
\usepackage{array}
\usepackage[caption=false,font=normalsize,labelfont=sf,textfont=sf]{subfig}
\usepackage{textcomp}
\usepackage{stfloats}
\usepackage{url}
\usepackage{verbatim}
\usepackage{graphicx}
\usepackage{cite}
\usepackage{amssymb}
\usepackage{physics}
\usepackage{epstopdf}
\usepackage{cases}
\usepackage{indentfirst}

\newtheorem*{Proof}{Proof:}
\theoremseparator{:}

\newtheorem{lemma}{Lemma}
\newtheorem{remark}{Remark}
\begin{document}

\title{Radio Map-Based Spectrum Sharing for Joint Communication and Sensing}

\author{Xinran~Fang, Wei~Feng, Senior Member, IEEE, Yunfei~Chen, Senior Member, IEEE, Dingxi~Yang, Ning~Ge, Zhiyong Feng, Senior Member, IEEE, \\ and Yue~Gao, Fellow, IEEE
	\thanks{X.~Fang, W.~Feng, Dingxi~Yang, and N. Ge  are with Department of Electronic Engineering, Tsinghua University, Beijing 100084, China.\\
	Y. Chen is with the School of Engineering, University of Warwick, Coventry CV4 7AL, U.K.  \\
    Z. Feng is with the School of Information and Communication Engineering, Beijing University of Posts and Telecommunications, Beijing 100084, China.\\
    Y. Gao is with School of Computer Science, Fudan University, Shanghai 200433, China. \\
	Corresponding author: Wei~Feng (e-mail: {fengwei}@tsinghua.edu.cn)}}

\maketitle

\begin{abstract}
The sixth-generation (6G) network is expected to provide both communication and sensing (C\&S) services. However, spectrum scarcity poses a major challenge to the harmonious coexistence of C\&S systems. Without effective cooperation, the interference resulting from spectrum sharing impairs the performance of both systems. This paper addresses C\&S interference within a distributed network. Different from traditional schemes that require pilot-based high-frequency interactions between C\&S systems, we introduce a third party named the radio map to provide the large-scale channel state information (CSI).  With large-scale CSI, we optimize the transmit power of C\&S systems to maximize the signal-to-interference-plus-noise ratio (SINR) for the radar detection, while meeting the ergodic rate requirement of the interfered user. Given the non-convexity of both the objective and constraint, we employ the techniques of auxiliary-function-based scaling and fractional programming for simplification. Subsequently, we propose an iterative algorithm to solve this problem. Simulation results corroborate our idea that the extrinsic information, i.e., positions and surroundings, is effective to decouple C\&S interference.
\end{abstract}



\section{INTRODUCTION}

The advent of the fifth-generation (5G) network has ushered in a myriad of applications, ranging from automatic driving and smart manufacturing to telemedicine. These applications not only demand high communication data rates but also require  precise sensing. The vision of enabling joint communication and sensing in the sixth-generation (6G) network has become increasingly clear. Driven by the ever-lasting demands, both communication and sensing (C\&S) systems are increasing their scale in full swing. However, their progress faces a significant obstacle of spectrum scarcity. The interference caused by spectrum sharing impairs the functionalities of both C\&S systems  \cite{f7,f8}. To coexist C\&S systems and further build an integrated C\&S network, the joint communication and sensing (JCAS) has become a heated topic in both academia and industry.

In the literature, investigations on JCAS can be classified into two categories \cite{f1}. The first is C\&S coexistence. In this case,  C\&S functions are implemented using separate hardware modules. Due to the spectrum scarcity, C\&S systems share spectrum resources. Consequently, they need to negotiate spectrum usage, power allocation, and temporal coordination to minimize interference.  Additionally, the similar hardware structure and signal processing procedures of wireless communication and radio sensing  have spurred the C\&S integration. In this case, C\&S functions are implemented by a common hardware. This integrative approach offers potential advantages in terms of cost-efficiency, reduced power consumption, and compactness and simultaneously facilitates a more versatile usage of radio resources. In the subsequent sections, we review related studies on C\&S integration, followed by discussions on C\&S coexistence.

Regarding C\&S integration, corresponding studies can be subdivided into three categories. The first is the radar-centric approach, that uses radar platforms to transmit information \cite{ff1,ff2,ff3,ff4}. In the early stage, Hassanien \emph{et al.} conducted several representative works, which modulated the radar sidelobe to achieve communication capabilities \cite{ff1}\cite{ff2}. However, such schemes have a drawback of low data rates. Furthermore, the index modulation was incorporated into radar-centric designs. This technique uses a set of candidates, such as waveforms \cite{ff3} and transmit codes \cite{ff4}, and embeds the information into the permutation and combination of these elements. Due to the vast number of possible permutations and combinations, the system can transmit data at Mbps rates without substantial modification of the primary radar signal. The second  is the communication-centric approach.  Related studies mainly investigated signal processing schemes to extract information from communication signals.  For example,  Kumari \emph{et al.} applied the preamble of IEEE 802.11ad for the target detection and parameter estimation, demonstrating that centimeter-level accuracy is achievable thanks to the perfect auto-correlation property of the preamble \cite{ff5}. Pucci \emph{et al.} explored the sensing capabilities of the orthogonal frequency division multiplexing (OFDM) waveforms. The authors derived the expressions of the  detection probability and the root mean squared error of the position and velocity estimation \cite{ff6}.
The third category is the novel waveform design, which aims to provide flexible C\&S services using a new dual-function waveform. For example, Sturm \emph{et al.} discussed the basic requirements of the dual-function waveform and the processing techniques to extract information \cite{ff8}. Liu \emph{et al.} proposed a solution to balance C\&S by minimizing a weighted objective of inter-user interference and deviations from constant-modulus waveforms. The authors also developed a low-complexity algorithm to facilitate practical implementations \cite{ff7}. Xiao \emph{et al.} proposed a full-duplex JCAS platform which leverages the idle window of probing pulses to transmit communication bits. The proposed scheme achieved a higher data rate than radar-centric schemes and better sensing performance than communication-centric schemes \cite{ff9}.

Compared with C\&S integration, which requires a long-term effort, the severe spectrum scarcity makes C\&S coexistence a more urgent task at present. In this paper, we thus concentrate on the interference mitigation between C\&S systems.
Related works are reviewed as follows.
In the early stage, related studies were mainly unilateral which optimized either the radar side or the communication side \cite{ref1}--\cite{ref10}. On the radar side, Bica \emph{et al.} used mutual-information-based criteria to optimize the radar waveform  \cite{ref1}. Babaei \emph{et al.}  introduced a null-space precoder that confines radar signals within the null space of interference channels  \cite{ref2}. Shi \emph{et al.} considered the inaccuracy of channel state information (CSI) and proposed a power minimization scheme, with the C\&S requirement serving as the constraints \cite{ref3}. Kang \emph{et al.} optimized the radar beampattern to manage its radiating energy in spatial directions and spectral frequency bands \cite{ref4}.  Alaee-Kerahroodi \emph{et al.} proposed a cognitive radar waveform and validated its effectiveness using a custom-built prototype \cite{ref21}.
On the communication side, E. H. G. Yousif \emph{et al.} devised a linear constraint variance minimization beamforming scheme to control the interference leakage from the base station (BS) to radars  \cite{ref6}. Raymond \emph{et al.} utilized a spatio-temporal analytical method to quantify the instantaneous and average interference suffered by radar from neighboring BSs. Building upon this analysis, they introduced an adaptive BS power control strategy that simultaneously safeguards primary radar operations while ensuring effective cellular communication \cite{ad3}. Singh \emph{et al.} explored spectrum sharing between a full-duplex multi-input and multi-output (MIMO)  BS and a MIMO radar. The authors proposed a power allocation scheme which takes the detection probability as the objective, subject to the signal-to-interference-plus-noise ratio (SINR) requirements for both uplink and downlink communications \cite{ad1}. Labib \emph{et al.} examined the coexistence between the radar and LTE system. The authors proposed a stochastic optimization scheme to maximize the BS transmit power within the bound of acceptable interference at the radar side \cite{ad4}.
Liu \emph{et al.} investigated the coexistence between a MIMO radar and a multi-user MIMO communication system. They proposed a robust communication precoding scheme to maximize the detection probability while ensuring the minimal SINR of downlink users \cite{ref7}.  Moreover, the authors further paid attention to practical limitations in C\&S cooperation \cite{ref17}. They investigated the channel estimation and radar working mode judgment under different levels of prior knowledge of the radar waveform. Recently, Wang \emph{et al.} and Shtaiwi \emph{et al.} introduced the reconfigurable intelligent surface (RIS) to the C\&S network. Both of them investigated the joint BS and RIS precoding scheme to exploit the degrees of freedom (DoFs) of RIS for controlling C\&S interference \cite{ref9,ref10}.

Due to the limited DoFs, the unilateral schemes sometimes are  insufficient to control interference effectively. The joint C\&S cooperation is accepted as a more effective means  \cite{ref11}--\cite{ref13}. Wang \emph{et al.} considered  a single-antenna C\&S network  and proposed a joint power allocation scheme from both radar-centric and communication-centric perspectives \cite{ref11}. Furthermore, the authors extended the power allocation scheme to multi-carrier C\&S systems with additional consideration of the clutters \cite{ad2}. Martone \emph{et al.} proposed a multi-objective optimization scheme, which adjusts power, frequency, and bandwidth parameters to maximize both C\&S criteria \cite{ff10}. Considering the competing relationships between C\&S systems, Mishra \emph{et al.} modeled  C\&S interactions as a two-person zero-sum game and proposed a power allocation strategy to find the equilibrium between C\&S systems \cite{ff11}. Tian \emph{et al.} applied mutual information as the metric of radar sensing and proposed a power allocation scheme that simultaneously maximizes mutual information of radar sensing and communication data rate  \cite{ff12}.
Li \emph{et al.} investigated the matrix-completion-based radars in the coexistence  design. With additional DoFs of subsampling of this kind of radar, they established an optimization framework that maximized the radar SINR with the constraint of the communication rate \cite{ref12}. Rihan \emph{et al.} investigated the spectrum sharing between MIMO radar and MIMO communication systems. The authors proposed a two-tier alternating optimization scheme based on the interference alignment approach, which optimizes  transmit precoders and receive spatial filters to maximize both the radar SINR and communication sum rate \cite{123}.
Qian \emph{et al.} jointly optimized the radar transmit waveform, receive filter, and code book of the communication system in a clutter environment. To ensure the estimation accuracy, the authors took the waveform similarity as a constraint in the optimization \cite{ref14}. Given that the interference generated by pulse radars is not constant, Zheng \emph{et al.} proposed a novel metric named the compound rate to measure the communication performance under varying  interference \cite{ref15}. He \emph{et al.} proposed a robust coexistence scheme with the objective of minimizing C\&S power. The probability that C\&S SINRs are below a certain threshold was constrained within a small value \cite{ref116}. In addition to optimizing the detection performance, Cheng \emph{et al.} minimized the Cramér-Rao Bound of the angle estimation by optimizing the radar waveform and communication transmit vectors \cite{ref18}. Grossi \emph{et al.} maximized the energy efficiency of the communication system while ensuring the received SINR of radars \cite{ref19}. Qian \emph{et al.} jointly optimized the radar receiver and communication transmitter to maximize the mutual information between the probing signals and received echoes \cite{ref13}.

Although the studies mentioned above offered valuable insights into mitigating interference between C\&S systems, a gap persists between the theoretical investigations and practical deployment.  In many cases, practical conditions cannot support the perfect CSI assumption used in the literature. First, obtaining full CSI requires pilot-based channel estimation. It is nearly impossible for existing C\&S systems to alter their hardware and software for receiving the counterpart signals. In addition, acquiring full CSI is costly. C\&S systems have to send pilots, estimate CSI, and exchange CSI data during each channel coherence time, resulting in heavy signaling overhead.  Moreover, C\&S systems are not robust using the pilot-based cooperation. The inaccuracy of CSI would degrade cooperation efficiency, impacting the provisions of C\&S services. Given these challenges, the pilot-based C\&S cooperation maybe not  a practical choice in real-world applications.

Motivated by the above drawbacks,  we propose a radio-map-based C\&S cooperation framework. A third party named the radio map is introduced to exploit extrinsic environmental information to estimate the large-scale CSI.
By doing so, the pilot-based high-frequency C\&S interactions are bypassed. The main contributions of this paper are summarized as follows:
\begin{enumerate}
	\item We propose a radio-map-based C\&S cooperation framework to mitigate C\&S  interference in a distributed network. Unlike traditional channel estimation relying on pilots, the radio map utilizes extrinsic environmental information to estimate the large-scale CSI. Leveraging the large-scale CSI, we propose a joint power allocation scheme that takes the radar SINR as the objective and the ergodic user rate as the constraint.
	\item  The formulated problem includes both the non-convex objective and constraint. To tackle the absence of a closed-form expression of the ergodic rate, we construct auxiliary functions and employ the scaling technique. To handle the fractional expression of the radar SINR, we apply the quadratic transformation. Building upon these techniques, we propose an iterative algorithm to solve this problem in low complexity.
	\item
	We employ a learning-based method to construct the radio map and utilize its predicted large-scale CSI to implement the proposed power allocation scheme. Simulation results validate the effectiveness of incorporating extrinsic information to coexist C\&S systems in a loosely cooperated manner.
\end{enumerate}

In addition, this work represents an extension of our previous work \cite{our}. Unlike \cite{our}, which assumed the radio map is ready-made, this work unites the map construction and usage to provide a complete radio-map-based framework. We enhance the mathematical work with detailed proofs and conduct unified simulations by incorporating the radio map and proposed power allocation scheme.
More in-depth discussions and insights are also provided throughout the paper. Therefore, this work is a necessary extension and enhancement of our previous work.

The remainder of this paper is organized as follows. In Section \ref{sec2}, we introduce the system model and concrete the concept of the radio map. Section \ref{sec3} presents the joint power allocation scheme and the iterative algorithm. Section \ref{sec4} introduces simulation results and Section \ref{sec5} draws conclusions.

Throughout this paper, vectors and matrices are represented by lower and upper boldface symbols, respectively. $\mathbb{C}^{M \times N}$ represents the set of $M\times N$ complex matrices and $(\cdot)^H$ is the operation of the transpose conjugate. The complex Gaussian distribution of zero mean and $\sigma^2$ variance is denoted as $\mathcal{CN}(0,\sigma^2)$. $\mbox{E}_s (\cdot)$ is the expectation operation with respect to $s$.

\section{SYSTEM MODEL}
\label{sec2}

\begin{figure}[t]
	\centering	\includegraphics[width=1\linewidth]{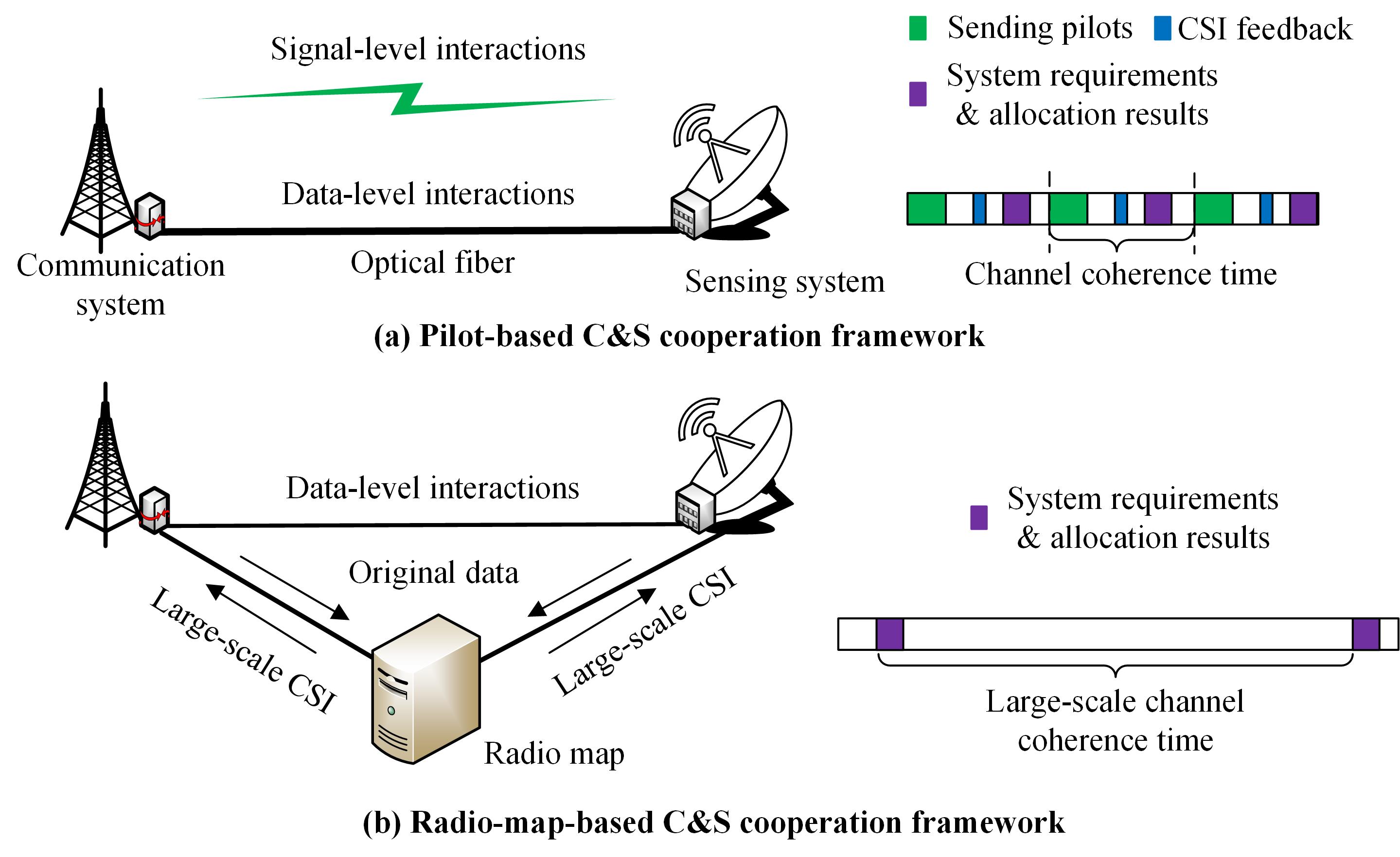}
	\caption{Illustration of the pilot-based C\&S cooperation framework and the proposed radio-map-based C\&S cooperation framework. }
	\label{Fig.3}
\end{figure}

In Fig. \ref{Fig.3}, we illustrate the pilot-based C\&S cooperation framework and the proposed radio-map-based C\&S cooperation framework.   As shown in Fig. \ref{Fig.3} (a), the pilot-based C\&S cooperation framework requires both signal-level and data-level interactions between C\&S systems. Within each channel coherence time, C\&S systems exchange pilots, CSI data, C\&S requirements, and allocation results. The two systems are tightly coupled by high-frequency and heavyweight interactions. In contrast,  the radio-map-based framework only requires low-frequency and lightweight C\&S interactions. Instead of using pilots to estimate channel conditions, C\&S systems send original data to the radio map and get the large-scale CSI back. With the radio-map-predicted large-scale CSI, C\&S systems only need to negotiate their requirements and resource allocation results. Their negotiation frequency is the same as the varying frequency of the large-scale fading, which changes slowly. C\&S systems thus work in a loosely cooperated manner.

\begin{figure}[t]
	\centering	\includegraphics[width=\linewidth]{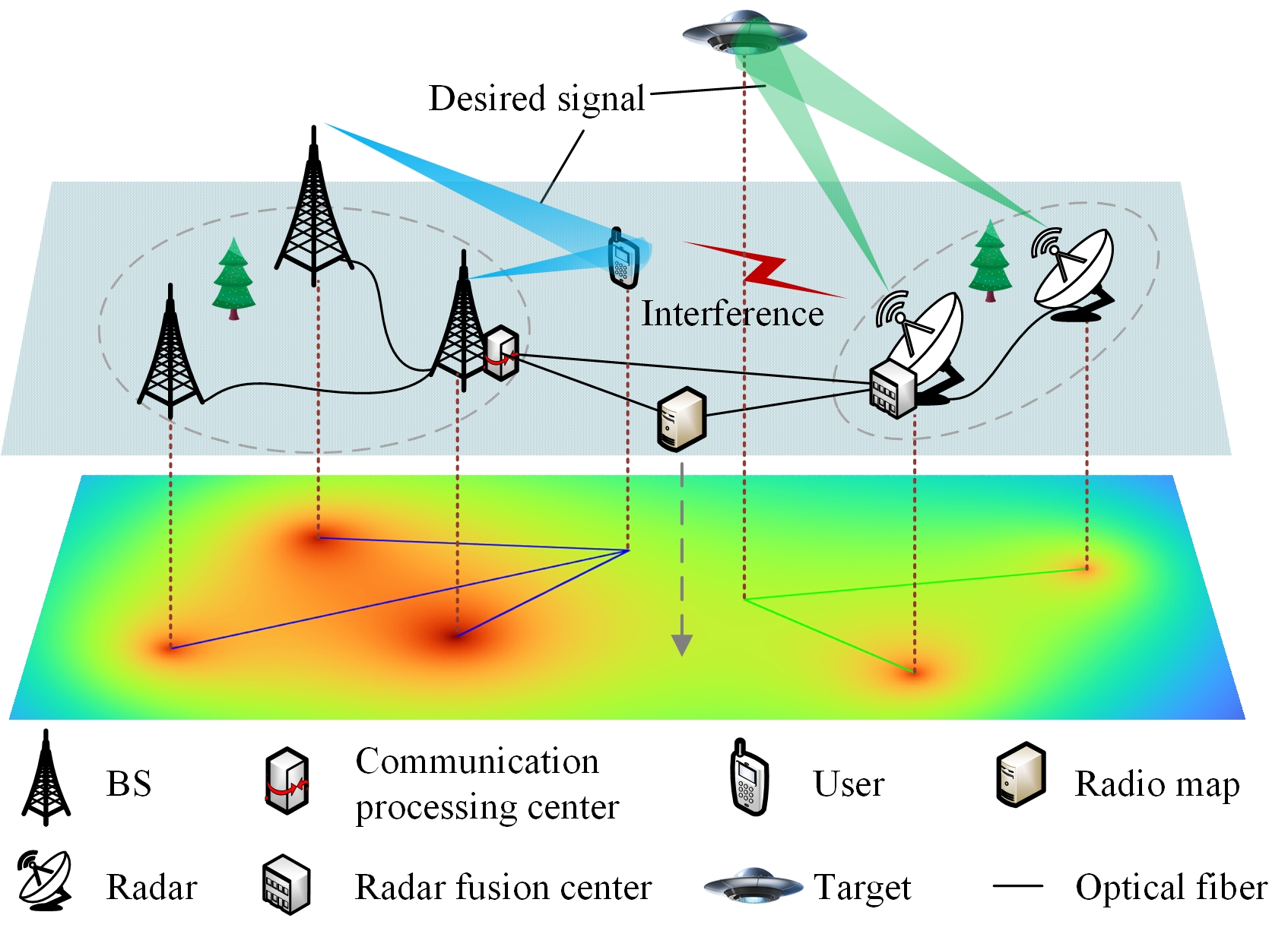}
	\caption{Illustration of a distributed network consisting of distributed C\&S systems. A third party named the radio map is introduced to provide the large-scale CSI.}
	\label{Fig.1}
\end{figure}

Based on the proposed framework, we illustrate our radio-map-based distributed C\&S network in Fig. \ref{Fig.1}. The communication system is composed of $M_c$ distributed BSs. These BSs are connected to a common processing center. They serve users in a user-centric manner. To avoid inter-user interference, different users are allocated with orthogonal frequency bands.  The sensing system is composed of $M_r$ distributed radars. These radars operate in the same frequency band and use the orthogonal signals to tell from each other. Radars detect a common target. The detection results are sent to the fusion center to calculate the final decision. In addition to C\&S systems, a third party named radio map is placed in the network to estimate the large-scale CSI. To facilitate the negotiation, optical fibers connect the communication processing center, radar fusion center, and radio map. In this network, C\&S systems share spectrum resources. We concentrate on one user whose allocated frequency band is overlapped with radars. The models and expressions can be extended to the case of multiple interfered users with few changes.

%The radio map collects original data from C\&S systems and provides the large-scale-CSI map in return.

\subsection{RADIO MAP MODEL}
\label{sec2a}

\begin{figure}[t]
	\centering	\includegraphics[width=1\linewidth]{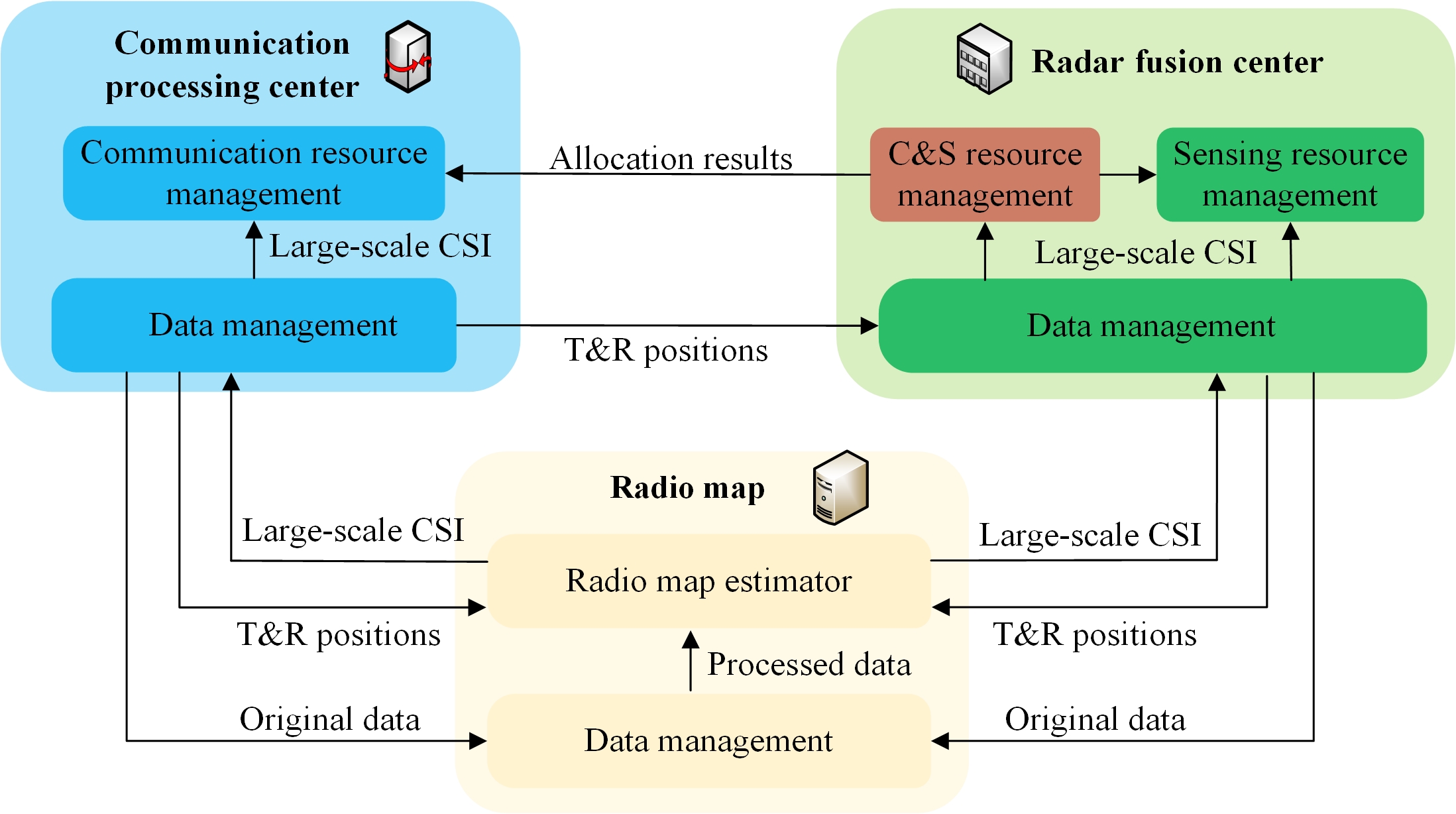}
	\caption{Illustration of the working regime of the proposed radio-map-based C\&S cooperation framework. The modules of the communication system, sensing system, and the radio map are depicted by the blue, green, and yellow blocks, respectively.}
	\label{Fig.2}
\end{figure}

In Fig. \ref{Fig.2}, we illustrate the working regime of the radio-map-based C\&S cooperation framework. As depicted by the yellow blocks, the radio map includes two key modules: data management and radio map estimator. The practical implementation of the radio map involves three steps: data collection and map construction, integration with C\&S systems, and map update. Initially, the data management module extensively collects data.  These data are collected by the approaches such as numerical calculations using the ray-tracing method and on-site measurements \cite{233}. The original data generated by C\&S systems in their daily operations can also be sent to the radio map. The data are preprocessed by the data management module and handed over to the radio map estimator, which is responsible for constructing the radio map. In the literature, the map construction methods can be divided into two categories: model-based and data-based.  Model-based methods are based on a prior model and use collected data to estimate model parameters. Conventional channel models are a kind of model-based radio maps.
These channel models are broadly categorized according to the transmission environment, e.g., urban, sub-urban, rural, and sub-rural \cite{r2-3}, and the model parameters are estimated using the data collected from corresponding environments. However, due to the rough division, radio maps generated from model-based methods cannot accurately describe specific environments. Data-based methods, on the other hand, use data exclusively to construct the radio map.  Related data-learning methods include kernel-based learning, kriging, dictionary learning, and deep learning \cite{r2-1}. These radio maps are environmental-aware and capable of providing accurate radio information for the areas where the data are collected. However, data-based methods suffer from the high cost of data management and high computational complexity \cite{r2-5}.
Therefore, some studies begin to explore combining these two methods in the construction \cite{r2-2}\cite{r2-4} \footnote{In this paper, the radio map exclusively refers to the map that is constructed for a certain area and is environmental-aware. }.
Afterwards, the second step is to integrate the radio map into C\&S systems. The radio map provider needs to negotiate with C\&S systems on the specifics such as the required original data, the provided information and the frequency of this provision.
The third step involves the map update. To reflect the dynamic environment, the data management module persistently gathers and processes data. The newly collected data are periodically relayed to the radio map estimator, which updates the radio map to capture the dynamics of the real world.

In our framework, the radio map works to provide the large-scale CSI. It maps position data, i.e., the transmitting and receiving (T\&R) positions, into large-scale CSI \cite{add}. This position information is collected by the data management modules of C\&S systems. They send T\&R positions to the radio map and get the corresponding large-scale CSI back. With the large-scale CSI, the C\&S resource management module runs the built-in algorithm, calculating the resource allocation results. Then, these results are sent to resource management modules of C\&S systems and being implemented. In the presented figure, the C\&S resource management module resides in the radar side. Consequently, the communication system sends its T\&R positions to the radar fusion center, enabling the C\&S resource management module to know the large-scale CSI of both systems.  The C\&S resource management module can alternatively be positioned within the communication processing center, integrated into the radio map, or function as an independent party to provide interference mitigation solutions. The specific configuration can be flexibly adapted to practical situations.

%{\color{blue}Moving forward, we discuss the applicability of the proposed framework under different moving-speed conditions. From Fig. \ref{Fig.2} we can see that the update frequency of the large-scale CSI determines the interaction frequency among the radio map and C\&S systems. In relative static or slow-moving scenario, that the large-scale CSI is slow-varying, the radio map and C\&S systems only need low-frequency interactions to update the large-scale CSI and implement the C\&S coordination scheme. This saves C\&S systems from high-frequency single-level interactions compared with the pilot-based scheme.  While in high-speed scenarios, the proposed framework also needs high-frequency interactions to update the position information and get the corresponding large-scale CSI. The superiority of the proposed scheme compared with traditional pilot-based schemes diminishes. Therefore, the proposed framework is more applicable for the slow-moving scenarios.} %Therefore, the superiority of our proposed framework is more appliable to the

%\begin{figure}[htbp]
%	\centering	\includegraphics[width=0.8\linewidth]{MLP.jpg}
%	\caption{\color{blue}Illustration of the MLP structure. The network consists of a two-dimensional input layer, one-dimensional output layer, and five hidden layers with  [32,64,128,64,32] neurons. }
%	\label{r2-2}
%\end{figure}

\begin{figure*}[htbp]
	\centering	\includegraphics[width=0.8\linewidth]{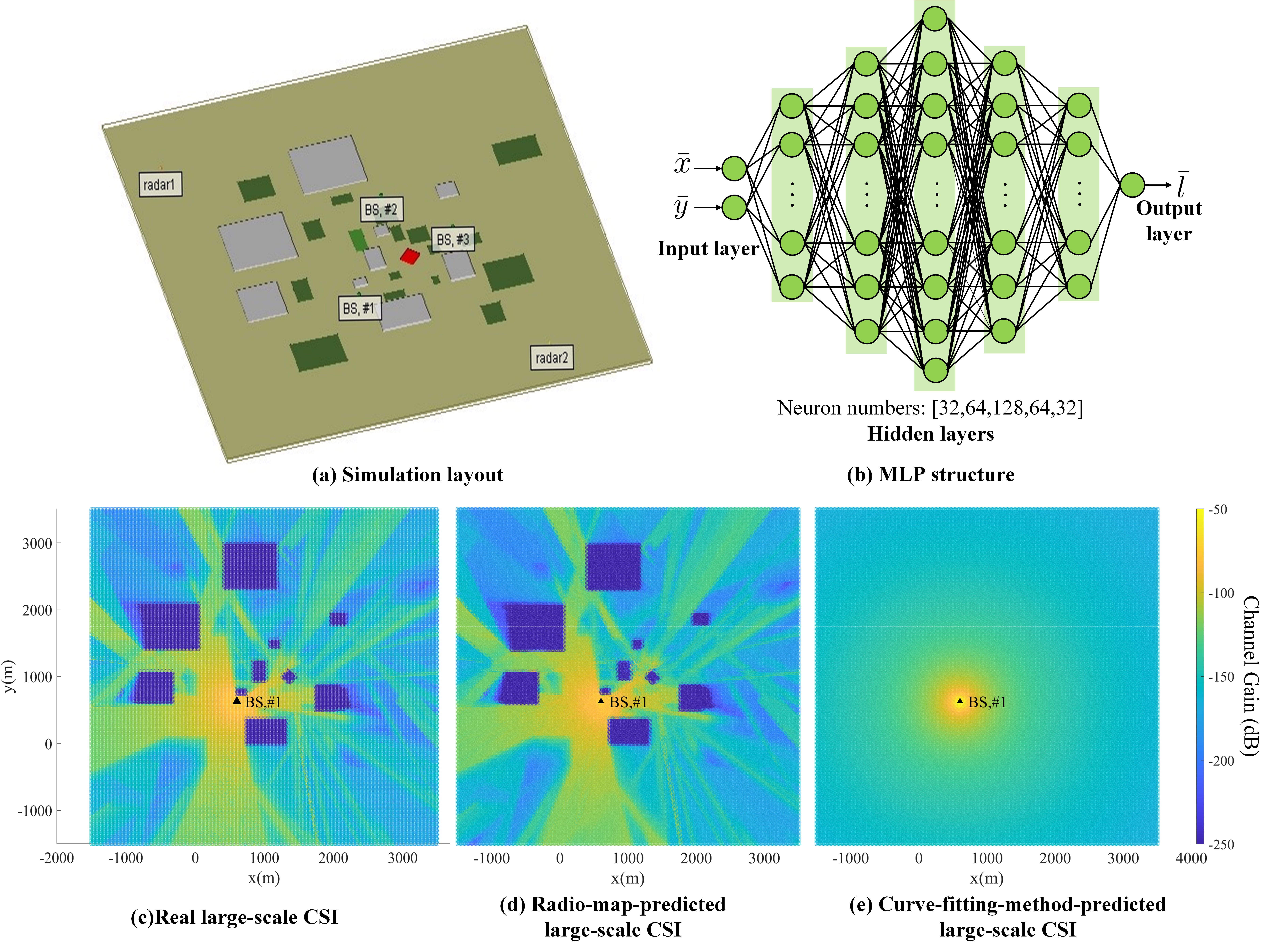}
	\caption{Illustration of the large-scale CSI maps generated by Wireless Insite, the radio map and the curve fitting method. (a) Simulation layout, (b) MLP structure
	(c) Real large-scale CSI calculated by Wireless Insite, (d) Radio-map-predicted large-scale CSI, and (e) Curve-fitting-method-predicted large-scale CSI. The MLP consists of a two-dimensional input layer, one-dimensional output layer, and five hidden layers with [32,64,128,64,32] neurons.
	The large-scale CSI maps include one million data points each. These data are obtained by dividing the layout into one million grid points and calculating the channel gain between each grid point and the BS, $\#1$, using the Wireless Insite, radio map and curve fitting method.}
	\label{fig1}
\end{figure*}

To demonstrate the feasibility of the proposed framework, we present a case study of the radio map. A ray tracing software named Wireless Insite was introduced to calculate the large-scale CSI, which is regarded as the ground truth data.  In Fig. \ref{fig1} (a), we illustrate the layout set in the software. It is a $5000$ m $\times\ 5000$ m rectangular area,  covering the range of $[-1500 \ \text{m}, 3500 \ \text{m}]$. There are randomly distributed buildings and lawns. In this simulation, we equally divided the layout into $4\times10^4$ grids and placed a virtual receiver at each grid point. The software was used to calculate the channel gain between the transmitters (BSs and radars) and virtual receivers. We thereby obtained $4\times10^4$  data points for each BS and radar.
Using the $4\times10^4$ data points and the corresponding positions of the virtual receiver, we trained a multi-layer perceptron (MLP) for each BS and radar. The structure of the MLP is presented in Fig. \ref{fig1} (b). During data preprocessing, we applied the max-min normalization to scale the data within the range of $[-1, 1]$. Our MLP takes the normalized receiving position, i.e, $[\bar{x},\bar{y}]$, as the input and takes the normalized channel gain, i.e., $\bar{l}$, as the output. The MLP has five hidden layers with $[32,64,128,64,32]$ neurons and between these layers the hyperbolic tangent function is used as the activation function.  For training, we employed the Adaptive Moment Estimation (ADAM) optimizer and used the default parameters as defined in TensorFlow: learning rate of $0.001$, $\beta =(0.9, 0.999)$, and $\epsilon=1e-8$ \cite{tensor}.  The network was trained over $6000$ epochs with a batch size of $128$, and the mean square error (MSE) was the loss function. Additionally, the learning rate was halved every $2000$ epochs to refine the learning process. In Fig. \ref{fig1} (c), we present the large-scale CSI map estimated by the radio map estimator of BS, $\#1$. For comparison, large-scale CSI maps generated by Wireless Insite and estimated by the curve fitting method are also presented. The curve fitting method is based on a generic channel model \cite{232}:
\begin{equation*}
	L(d,f_c)=\alpha \log_{10}(d)+20\log_{10}(f_c)+\beta,
\end{equation*}
where $L$ represents the path loss, $d$ is the T\&R distance, and $f_c$ is the carrier frequency. We applied the least squares method to calculate $\alpha$ and $\beta$. The used data
are the same as those used for constructing the radio map.

Through comparisons, we can see that the radio map provides a more accurate approximation to the real large-scale CSI compared to the curve fitting method. This is because the radio map is environment-aware. By incorporating  T\&R positions, the radio map considers the losses caused by specific geographical factors along this link. It thus delivers more accurate large-scale CSI than generic empirical models.
In addition, we discuss the applicability of the proposed framework for different moving-speed targets. As shown in Fig. \ref{Fig.2}, the update frequency of the large-scale CSI determines the interaction frequency among the radio map and C\&S systems. In the scenario of slow-moving targets, the large-scale CSI is slow-varying. As a result, the radio map and C\&S systems only need low-frequency interactions.  While in the contexts with high-speed targets,
C\&S systems need high-frequency interactions to update the target position and large-scale CSI. This lets C\&S systems to be tightly coupled. In addition, to adequately cover the area of target activities, the radio map needs to be large, which greatly increases its construction and maintenance costs. The benefits of our radio-map-based framework over traditional pilot-based framework become less pronounced. Therefore, this paper primarily focuses on the applicability of our radio-map-based framework in scenarios with slow-moving targets. Potential extension to fast-moving targets necessitates the development of cost-effective algorithms for large-scale radio map construction and update.

Moreover, the radio map, recording the information of the electromagnetic environment, its usage is more than C\&S coordination. With enhanced environmental awareness, the radio map contributes to the efficiency of networks, facilitating a smarter utilization of the  spectrum resources. For example, radio maps help in dynamic spectrum management by identifying spectrum usage patterns and detecting interference sources. They also assist in urban sensing by providing electromagnetic profile data, which help determine the location and size of buildings. Additionally, radio maps play an active role in localization, enabling accurate positioning by comparing real-time signal measurements with pre-constructed radio maps. In security applications, radio maps assist in detecting unauthorized devices and monitoring signal activities within a specified area. Radio maps can also be applied in environmental monitoring, such as assessing air conditions by providing signal propagation attenuation data. By incorporating the radio map with other advanced technologies, e.g., machine learning and Internet of things (IoT) \cite{feng}, the sensing accuracy, communication efficiency, and overall intelligence will be greatly enhanced. However, realizing the full potential of radio maps necessitates dedicated research efforts in several key areas. Fast and efficient map construction and update are critical to ensure the radio map remain effective for the dynamic electromagnetic environment. Seamless integration with C\&S systems requires the development of interoperable standards and protocols. Moreover, striking a balance between the costs and benefits of the radio map is paramount to ensure its widespread  deployment. We call for more research efforts to explore these promising avenues further.

\subsection{CHANNEL MODEL}
In this subsection, we introduce the channel model. We assume that BSs and radars are equipped with a single antenna. This is because, in MIMO systems, the precoding scheme that exploits the multiplexing gain requires perfect and instantaneous CSI \cite{234}. With only large-scale CSI, multiple antennas are equivalent to a combined antenna for both BSs and radars. The interfered user is assumed to have $N_c$ antennas, which are used to receive signals from distributed BSs. On this basis, we denote the BS-user link and interference links, i.e., BS-radar link and radar-user link as $\mathbf{H}_{c}^H=[\mathbf{h}_{c_1}^H,\mathbf{h}_{c_2}^H,...,\mathbf{h}_{c_{M_c}}^H]\in\mathbb{C}^{N_c\times M_c}$, $\mathbf{h}_{c\rightarrow r_i}\in\mathbb{C}^{M_c\times 1}$, and  $\mathbf{h}_{r_i \rightarrow c}\in\mathbb{C}^{1\times N_c}$, respectively, where $i$ denotes the index number of radars. They are modeled by the composite channel model:
\begin{equation}
	\begin{aligned}
		\mathbf{H}_{c}&=(\mathbf{L}_{c})^{1/2}\mathbf{S}_{c},\\ \mathbf{h}_{c\rightarrow r_i}&=(\mathbf{L}_{c \rightarrow r_i})^{1/2}\mathbf{s}_{c\rightarrow r_i}, \\
		\mathbf{h}_{r_i\rightarrow c}&=(l_{r_i \rightarrow c})^{1/2}\mathbf{s}_{r_i\rightarrow c},
	\end{aligned}
\end{equation}
where $\mathbf{L}_c\in\mathbb{C}^{M_c\times M_c}$, $\mathbf{L}_{c\rightarrow r_i}\in\mathbb{C}^{M_c\times M_c}$, and $l_{r_i\rightarrow c}$ denote the large-scale fading. $\mathbf{L}_c$ and $\mathbf{L}_{c\rightarrow r_i}$ are diagonal matrices whose $j$-th diagonal elements are denoted as $l_{c_j}$ and $l_{c_j\rightarrow r_i}$.
$\mathbf{S}_c\in\mathbb{C}^{M_c\times N_c}$, $\mathbf{s}_{c\rightarrow r_i}\in\mathbb{C}^{M_c\times 1}$, and $\mathbf{s}_{r_i\rightarrow c}\in\mathbb{C}^{1\times N_c}$ denote the small-scale fading, whose elements are uniformly and independently conformed to the Gaussian distribution, i.e., $\mathcal{CN}(0,1)$. For the two-way links between the radars and  target, i.e., $\mathbf{H}_r=[\mathbf{h}_{r_1},....,\mathbf{h}_{r_{M_r}}]\in\mathbb{C}^{M_r\times M_r}$, they are modeled by the determined channel model. This is because when we consider the target in the sky, the corresponding air-to-ground channel is dominated by the line-of-sight path with little random fading.%Thereby, we assume the radio map estimates the slow-varying and definite CSI, i.e.,   $\mathbf{L}_{[\cdot]}$ and $\mathbf{H}_r$. The fast-varying small-scale fading, i.e., $\mathbf{S}_{[\cdot]}/\mathbf{s}_{[\cdot]}$ is unknown by the network.
%Without loss of generality, we assume that the small-scale fading conforms the $\mathcal{CN}(0,1)$ distribution.

\subsection{SIGNAL MODEL}
\begin{figure*}
		\begin{equation}
			\label{add1}
			\begin{aligned}
				y_{r_i}(t)=&\sum_{j=1}^{M_r}h^H_{r_j\rightarrow r_i}(t)\sqrt{p_{r_j}}\bigg(\sum_{n=0}^{N-1}u_{r_j}(t-\tau_{r_j\rightarrow r_i}-n\tau)\bigg)e^{j2\pi(f_c+f_{d_{r_j\rightarrow r_i}})(t-\tau_{r_j\rightarrow r_i})}\\
				+&\sum\limits_{j=1}^{M_c}h^H_{c_j\rightarrow r_i}(t)\sqrt{p_{c_j}}\bigg(\sum\limits_{n=0}^{N-1}u_{c_j}(t-\tau_{c_j\rightarrow r_i}-n\tau)\bigg)e^{j2\pi f_c(t-\tau_{c_j\rightarrow r_i})}+w_{r_i}(t),
			\end{aligned}
		\end{equation}

		\begin{equation}
			\label{ad1}
			\begin{aligned}
				y_{r_i}(t)=&\alpha_{i,i} h^H_{r_i\rightarrow r_i}(t)\sqrt{p_{r_i}}{\chi}_{i,i}(t-\tau_{r_i\rightarrow r_i}-n\tau,\Delta^{f_d}_{r_i\rightarrow r_i})e^{j2\pi \tilde{f}_{{d_{r_i\rightarrow r_i}}}t}\\
				+&\sum_{j=1,j\neq i}^{M_r}\alpha_{j,i}h^H_{r_j\rightarrow r_i}(t)\sqrt{p_{r_j}}{\chi}_{j,i}(t-\tau_{r_j\rightarrow r_i}-n\tau,\Delta^{f_d}_{r_j\rightarrow r_i})e^{j2\pi \tilde{f}_{{d_{r_j\rightarrow r_i}}}t} \\
				+&\sum_{j=1}^{M_c}h^H_{c_j\rightarrow r_i}(t)\sqrt{p_{c_j}}e_j(t-\tau_{c_j\rightarrow r_i}-n\tau)+w_{r_i}(t),
			\end{aligned}
		\end{equation}
		\begin{equation}
			\label{add3}
			\begin{aligned}
				y_{r_i}(n)=&\alpha_{i,i} h^H_{r_i\rightarrow r_i}(t^r_{i,n})\sqrt{p_{r_i}}{\chi}_{i,i}(\Delta^{\tau}_{r_i\rightarrow r_i} ,\Delta^{f_d}_{r_i\rightarrow r_i})e^{j2\pi \tilde{f}_{{d_{r_i\rightarrow r_i}}}t^r_{i,n}}\\
				+&\sum_{j=1,j\neq i}^{M_r}\alpha_{j,i}h^H_{r_j\rightarrow r_i}(t^r_{i,n})\sqrt{p_{r_j}}{\chi}_{j,i}(\Delta^{\tau}_{r_j\rightarrow r_i},\Delta^{f_d}_{r_j\rightarrow r_i})e^{j2\pi \tilde{f}_{{d_{r_j\rightarrow r_i}}}t^r_{i,n}}\\
				+&\sum_{j=1}^{M_c}h^H_{c_j\rightarrow r_i}(t^r_{i,n})\sqrt{p_{c_j}}e_j(\Delta^{\tau}_{c_j\rightarrow r_i})+w_{r_i}(t^r_{i,n}), n=1,...,N,
			\end{aligned}
		\end{equation}
		\begin{small}
			\begin{equation}
				\label{1}
				\begin{aligned}
					&\rho_i(\mathbf{P}_r,\mathbf{P}_c)\\
					&=\frac{p_{r_i}|h_{r_i\rightarrow r_i}(t^r_{i,n})|^2|\chi_{i,i}(\Delta^{\tau}_{r_i\rightarrow r_i},\Delta^{f_d}_{r_i\rightarrow r_i})|^2}{\sum\limits_{j=1, j\neq i}^{M_r}p_{r_j}|h_{r_j\rightarrow r_i}(t^r_{i,n})|^2|\chi_{j,i}(\Delta^{\tau}_{r_j\rightarrow r_i},\Delta^{f_d}_{r_j\rightarrow r_i})|^2+\mathbb{E}\bigg(\sum\limits_{j=1}^{M_c}p_{c_j}e_j(\Delta^{\tau}_{c_j\rightarrow r_i})^He_j(\Delta^{\tau}_{c_j\rightarrow r_i})|h_{c_j\rightarrow r_i}(t^r_{i,n})|^2+w_{r_i}^H(t^r_{i,n})w_{r_i}(t^r_{i,n})\bigg)}\\
					&=\frac{p_{r_i}|h_{r_i\rightarrow r_i}|^2|\chi_{i,i}(\Delta^{\tau}_{r_i\rightarrow r_i}\Delta^{f_d}_{r_i\rightarrow r_i})|^2}{\sum\limits_{j=1, j\neq i}^{M_r}p_{r_j}|h_{r_j\rightarrow r_i}|^2|\chi_{j,i}(\Delta^{\tau}_{r_j\rightarrow r_i},\Delta^{f_d}_{r_j\rightarrow r_i})|^2+\sum\limits_{j=1}^{M_c}l_{c_j\rightarrow r_i}p_{c_j}+\sigma^2},
				\end{aligned}
			\end{equation}
		\end{small}
\hrule
\end{figure*}
Denote the transmit signals of radars and BSs  as $\mathbf{x}_r(t)=[x_{r_1}(t),...,x_{r_{M_r}}(t)]^T$ and $\mathbf{x}_c(t)=[x_{c_1}(t),...,x_{c_{M_c}}(t)]^T$, respectively. The radar emits $N$ pulses within a coherent processing interval (CPI): $N$ repetitions of an orthogonal waveform, i.e., $x_{r_i}(t)=\sum\limits_{n=0}^{N-1}\sqrt{p_{r_i}}u_{r_i}(t-n\tau)e^{j2\pi f_ct},\ t\in[0,\tau], n=1,..., N$, where $p_{r_i}$ is the transmit power of the $i$-th radar, $u_{r_i}(t)$ is the complex envelop of a radar pulse, which has unit energy, $f_c$ denotes the carrier frequency, and $\tau$ is the pulse duration. Similarly, the communication signal is expressed as: $x_{c_i}(t)=\sum\limits_{n=0}^{N-1}\sqrt{p_{c_i}}u_{c_i}(t-n\tau)e^{j2\pi f_ct}, \ t\in[0,\tau], n=1,..., N$, where $p_{c_i}$ denotes the transmit power of the $i$-th BS and $\tau$ is the duration of a communication symbol\footnote{In this paper, we assume that the pulse duration of radars is the same as the symbol duration of BSs and C\&S systems use the same sampling rate, i.e., $\frac{1}{\tau}$ \cite{123}\cite{ref19}.}. Denote the bistatic delay from the $j$-th radar to the $i$-th radar as $\tau_{r_j\rightarrow r_i}$, the bistatic delay from the $j$-th BS to the $i$-th radar as $\tau_{c_j\rightarrow r_i}$, and the Doppler shift from the $j$-th radar to the $i$-th radar as $f_{d_{r_j\rightarrow r_i}}$. The received signal of the $i$-th radar is given by \eqref{add1} (shown at the top of the next page), where $w_{r_i}(t)$ is the  noise received by the $i$-th radar.

The received signal first goes through the down conversion (remove the term $e^{j2\pi f_ct}$ in \eqref{add1}) and then goes through a matched filter (MF) to extract the echoes of its own probing signals. The MF of the $i$-th radar is $u_i^H(-t)e^{j2\pi \tilde{f}_{{d_{r_{i}\rightarrow r_i}}}t}$ \cite{ad5}, where $\tilde{f}_{{d_{r_{i}\rightarrow r_i}}}$ is the estimated Doppler Shift.  The output signal of the MF is given by \eqref{ad1} (shown at the top of the next page), where $\alpha_{j,i}\triangleq e^{-j2\pi (f_c+f_{d_{r_j\rightarrow r_i}})\tau_{r_j\rightarrow r_i}}$, $\Delta^{f_d}_{r_j\rightarrow r_i}\triangleq f_{{d_{r_{j}\rightarrow r_i}}}-\tilde{f}_{{d_{r_{j}\rightarrow r_i}}}$ is the Doppler shift error, $e_j(t-\tau_{c_j\rightarrow r_i}-n\tau)$ is the MF output of the $j$-th BS signal, and ${\chi}_{j,i}(t,f)$ is the ambiguity function, which is given by
\begin{equation}
	{\chi}_{j,i}(t,f)=\int u_{r_j}(v)u^H_{r_i}(v-t)e^{j2\pi fv}dv.
\end{equation}
Then, the output signal is sampled at the pulse rate, obtaining the samples at the time instants $t^r_{i,n}$, where $t^r_{i,n}=\tau_{r_i\rightarrow r_i}+\Delta^{\tau}_{r_i\rightarrow r_i}+n\tau, n=1,...,N$, and $\Delta^{\tau}_{r_i\rightarrow r_i}$ is the delay error. The output samples are given by \eqref{add3} (shown at the top of the next page), where $\Delta^{\tau}_{r_j\rightarrow r_i}\triangleq\tau_{r_i\rightarrow r_i}-\tau_{r_j\rightarrow r_i}$ and $\Delta^{\tau}_{c_j\rightarrow r_i}\triangleq\tau_{r_i\rightarrow r_i}-\tau_{c_j\rightarrow r_i}$. On this basis, we calculate the averaged SINR of the $i$-th radar, which is given by \eqref{1}\footnote{We assume the two-way channel between the radar and the target and the large-scale CSI of the interference channels keep unchanged within a CPI, and omit the time index $t^r_{i,n}$ in \eqref{1}.} (shown at the top of the next page), where  $\mathbf{P}_r=\text{diag}\{p_{r_1},...,p_{r_{M_r}}\}$, $\mathbf{P}_c=\text{diag}\{p_{c_1},...,p_{c_{M_c}}\}$,  $e_j(\Delta^{\tau}_{c_j\rightarrow r_i})$ is assumed to conform the $\mathcal{CN}(0,1)$ distribution and $w_{r_i}(t^r_{i,n})$ conforms the $\mathcal{CN}(0,\sigma^2)$ distribution.

To focus on the problem, we assume that the delay and Doppler shift are   compensated, i.e., $\Delta^{\tau}_{r_i\rightarrow r_i}=0, \Delta^{f_d}_{r_i\rightarrow r_i}=0$, and the orthogonality of radar waveforms holds across the whole delay-Doppler plane, i.e., ${\chi}_{i,i}(\Delta^{\tau}_{r_i\rightarrow r_i},\Delta^{f_d}_{r_i\rightarrow r_i})=1$ and ${\chi}_{j,i}(\Delta^{\tau}_{r_j\rightarrow r_i},\Delta^{f_d}_{r_j\rightarrow r_i})=0$. Under this assumption, the received SINR of the $i$-th radar is given by
\begin{small}
\begin{equation}
	\label{add4}
	\begin{aligned}
		&\rho_i(\mathbf{P}_r,\mathbf{P}_c)=\\
		&\frac{p_{r_i}|h_{r_i\rightarrow r_i}|^2|\chi_{i,i}(\Delta^{\tau}_{r_i\rightarrow r_i},\Delta^{f_d}_{r_i\rightarrow r_i})|^2}{\sum\limits_{j=1, j\neq i}^{M_r}p_{r_j}|h_{r_j\rightarrow r_i}|^2|\chi_{j,i}(\Delta^{\tau}_{r_j\rightarrow r_i},\Delta^{f_d}_{r_j\rightarrow r_i})|^2+\sum\limits_{j=1}^{M_c}l_{c_j\rightarrow r_i}p_{c_j}+\sigma^2}\\
		&=\frac{p_{r_i}|h_{r_i\rightarrow r_i}|^2}{\sum\limits_{j=1}^{M_c}l_{c_j\rightarrow r_i}p_{c_j}+\sigma^2}\triangleq\frac{p_{r_i}|h_{r_i\rightarrow r_i}|^2}{\sigma_r^2(\mathbf{P}_c)},
	\end{aligned}
\end{equation}
\end{small}
where $\sigma_r^2(\mathbf{P}_c)$ denotes the interference-plus-noise suffered by radar $i$.
In our setting, radar uses the generalized likelihood ratio test to judge the presence of the target. The detection probability and false alarm probability, denoted as $P_{D_i}$ and $P_{F_i}$, are calculated as follows \cite{ref3},
\begin{equation}
	\label{eq1}
	\left\{
	\begin{aligned}
		P_{D_i}(\rho_{i}(\mathbf{P}_r,\mathbf{P}_c),\mu_i)&=\big(1+\frac{\mu_i}{1-\mu_i}\frac{1}{1+N\rho_i(\mathbf{P}_r,\mathbf{P}_c)}\big)^{1-N} \\
		P_{F_i}(\mu_i)&=(1-\mu_i)^{N-1},
	\end{aligned}
	\right.
\end{equation}
where $\mu_i$ is the detection threshold determined by the false alarm probability. Given the requirement of the false alarm probability, $P_{D_i}$ is a function of the radar SINR. Therefore, we take the radar SINR as the sensing metric.

As for the communication system, since the distributed BSs work in a user-centric manner, they would estimate the transmission delay and compensate delay differences before sending communication signals. Therefore, the interfered user receives nearly aligned signals from distributed BSs, i.e., $\tau_c=\tau_{c_j}, \forall j$:
\begin{equation}
	\label{add5}
	\begin{aligned}
	&\mathbf{y}_c(t)=\sum\limits_{j=1}^{M_c}\mathbf{h}^H_{c_j}(t)\bigg(\sum\limits_{n=0}^{N-1}\sqrt{p_{c_j}}u_{c_j}(t-n\tau-\tau_{c})e^{j2\pi f_c(t-\tau_c)}\bigg)\\
	&+\sum_{j=1}^{M_r}\mathbf{h}^H_{r_j\rightarrow c}(t)\bigg(\sum\limits_{n=0}^{N-1}\sqrt{p_{r_j}}u_{r_j}(t-n\tau-\tau_{r_j})e^{j2\pi f_c(t-\tau_{r_j})}\bigg)\\
	&+\mathbf{w}_c(t),
	\end{aligned}
\end{equation}
where we do not consider the Doppler shift by assuming the interfered user is relatively static.
Afterwards, the communication signal goes through the down conversion and is sampled at symbol rate: $t^c_n=n\tau+\tau_c, n=1,...,N$. The sampled signal is given by
\begin{equation}
	\begin{aligned}
		&\mathbf{y}_c(n)=\sum\limits_{j=1}^{M_c}\mathbf{h}^H_{c_j}(t^c_n)\sqrt{p_{c_j}}u_{c_j}(0)e^{j2\pi f_c(t^c_n-\tau_c)}+\\
		&\sum_{j=1}^{M_r}\mathbf{h}^H_{r_j\rightarrow c}(t^c_n)\sqrt{p_{r_j}}u_{r_j}(\Delta^{\tau}_{r_j\rightarrow c})e^{j2\pi f_c(t^c_n-\tau_{r_j})}\\
		&+\mathbf{w}_c(t^c_n), n=1,...,N,
	\end{aligned}
\end{equation}
where $\Delta^{\tau}_{r_j\rightarrow c}\triangleq \tau_c-\tau_{r_j}$ and $\mathbf{w}_c(t^c_n)\in\mathbb{C}^{N_c\times 1 }$ is the white noise that conforms to the $\mathcal{CN}(0,\sigma^2\mathbf{I}_{N_c})$ distribution. The averaged interference-plus-noise suffered by the interfered user is calculated by
\begin{equation}
	\begin{aligned}
	\sigma_c^2(\mathbf{P}_r)=&\mathbb{E}\bigg(\sum_{j=1}^{M_r}p_{r_j}\mathbf{h}^H_{r_j\rightarrow c}(t^c_n)\mathbf{h}_{r_j\rightarrow c}(t^c_n)\times\\
	&u_{r_j}^H(\Delta^{\tau}_{r_j\rightarrow c})u_{r_j}(\Delta^{\tau}_{r_j\rightarrow c})+\mathbf{w}_c(t^c_n)\mathbf{w}_c^H(t^c_n)\bigg)\\
	=&\bigg(\sum\limits_{j=1}^{M_r}l_{r_j\rightarrow c}p_{r_j}+\sigma^2\bigg)\mathbf{I}_{N_c}.
	\end{aligned}
\end{equation}
Given the large-scale CSI, the ergodic rate of the interfered user, denoted as $\bar{R}(\mathbf{P}_c,\mathbf{P}_r)$, can be calculated:
\begin{equation}
	\label{R}
	\bar{R}(\mathbf{P}_c,\mathbf{P}_r)=\mathbb{E}_{\mathbf{S}_c}
	\big[\log_2\text{det}(\mathbf{I}_{N_c}+\frac{\mathbf{H}_c^H\mathbf{P}_c\mathbf{H}_c}{\sigma_c^2(\mathbf{P}_r)}\big)].
\end{equation}

\begin{remark}
	Since we assume the delay and Doppler shifts are perfectly compensated, \eqref{add4} and \eqref{R} give the optimistic description of the radar SINR and the communication ergodic rate. This aligns to the reality when the delay and Doppler shift errors are small and works effectively  for the literature investigation, e.g., \cite{ref11, ad2, ff12, 123}. Here, we  simply discuss how the delay and Doppler shift are measured and compensated in C\&S systems.

	As for the radar side, this measurement is implemented by applying a set of MFs with distinct Doppler shifts to match the received echoes and sampling the output at a given rate \cite{add4}. Therefore, the delay error and the Doppler shift error are given by: $|\Delta_{\tau}|\leqslant \frac{T_s}{2}, |\Delta_{f_d}|\leqslant \frac{\Delta f_d}{2}$, where $T_s$ is the sampling frequency and $\Delta_{f_d}$ is the frequency interval of the MFs.
	%This measurement is achieved by processing the echos using a fixed MF and sampling the MF output in fast- and small- times and then converting the samples to the frequency domain. %The delay and doppler error is determined by the fast and small time sampling rate, which is determined by the radar bandwidth and pulse repetition interval, respectively.
	As for the communication system, the user position is collected for the delay estimation. Based on the estimated delay, BSs compensate the delay differences in advance, ensuring the signals arrive at the same or closely matched time. Then, the receiver further mitigates time discrepancies according to the insert timing signals. The residual tiny differences would not impact the  symbol-level decoding. The Doppler shift compensation is achieved by signal processing methods such as the adaptive equalization. In terms of  fast-moving scenarios, 5G system uses the extended cycle prefix in OFDM waveforms to bypass the inter-symbol interference caused by the Doppler shift.

	Compared with the communication system that serves for the user who is cooperative, the radar needs to detect the unknown target, which makes it more difficult to perfectly compensate the delay and Doppler shift.
	In the case of no negligible error of the delay and Doppler shift,
	the auto-term in \eqref{1}  becomes less than 1 and the cross-term is no longer zero, i.e., $\chi_{i,i}(\Delta^{\tau}_{r_i\rightarrow r_i}\Delta^{f_d}_{r_i\rightarrow r_i})<1$ and $\chi_{j,i}(\Delta^{\tau}_{r_j\rightarrow r_i},\Delta^{f_d}_{r_j\rightarrow r_i})>0$, which lowers the value of the received SINR. We discuss the impacts of the inaccurate delay and Doppler shift compensation in the simulation.
	%In this paper, we do not delve into this issue and take it as an important investigation direction in the future. The impacts of the inaccurate delay and Doppler shift compensation are discussed in the simulation.
\end{remark}

\section{LARGE-SCALE-CSI-BASED POWER ALLOCATION}
\label{sec3}
\subsection{PROBLEM FORMULATION}
Leveraging the large-scale CSI, we take the minimal received SINR of radars as the objective and the ergodic rate requirement of the interfered user as the constraint.
Given the maximal transmit power of BSs and radars, denoted as $P_{cmax}$ and $P_{rmax}$, and the power budgets of C\&S systems, denoted as $P_{csum}$ and $P_{rsum}$, the joint power allocation problem is formulated as follows:
\begin{subequations}
	\label{P1}
	\begin{align}
		\mbox{(P1)} \max\limits_{\mathbf{P}_c,\mathbf{P}_r }\min\limits_{i\in[1,M_r]}\
		&\rho_i(\mathbf{P}_r, \mathbf{P}_c)   \\
		s.t. \ &\bar{R}(\mathbf{P}_c, \mathbf{P}_r) \geqslant R_{req} \label{3}\\
		&\mathbf{0}_{M_c} \preceq \mathbf{P}_c  \preceq P_{cmax}\cdot\mathbf{I}_{M_c} \label{4}\\
		&\mathbf{0}_{M_r} \preceq \mathbf{P}_r  \preceq P_{rmax}\cdot\mathbf{I}_{M_r} \label{5}\\
		&\operatorname{tr}(\mathbf{P}_c)\leqslant P_{csum} \label{6}\\
		&\operatorname{tr}(\mathbf{P}_r)\leqslant P_{rsum}, \label{7}
	\end{align}
\end{subequations}
where \eqref{3} is the ergodic rate requirement of the interfered user, \eqref{4} and \eqref{5} are the maximal transmit power constraints of BSs and radars, and \eqref{6} and \eqref{7} are the sum-power constraints of C\&S systems.  In addition, while the proposed scheme primarily takes the radar SINR as the objective, the communication performance is preferentially satisfied as we consider it in the constraint. In practical implementations, the ergodic rate threshold is adjusted to meet the requirement of the interfered user.
However, this needs manual adjustments of the threshold parameter to  provide satisfying C\&S solutions for different scenarios. To achieve a more adaptable balance between C\&S functions, a compound metric that can uniformly measure C\&S performance is necessary. We call for more research attention to this direction, which would present exciting revenues  for the further development of JCAS.

\subsection{ITERATIVE ALGORITHM}
\subsubsection{CLOSED-FORM APPROXIMATION}
(P1) is highly complex due to the non-convexity of the objective and  constraint. The most challenging aspect is the ergodic rate expression \eqref{R}, which is encapsulated within the expectation operation and lacks a closed-form representation.  To address this challenge, we first introduce an approximation to $\bar{R} (\mathbf{P}_c, \mathbf{P}_r)$ \cite{reff}:
\begin{equation}
	\label{11}
	\begin{aligned}
		\bar{R}_{ap}(\mathbf{P}_c,\mathbf{P}_r,v^*)=\sum\limits_{j=1}^{M_c}\log_2\big(1+\frac{N_cl_{c_j}p_{c_j}}{v^*\sigma_c^2(\mathbf{P}_r)}\big)\\
		+N_c\log_2(v^*)-N_c\log_2(e)[1-\frac{1}{v^*}],
	\end{aligned}
\end{equation}
where $v^*$ is determined by the following fixed-point equation,
\begin{equation}
	\label{10}
	1-\frac{1}{v^*}=\sum\limits_{j=1}^{M_c}\frac{l_{c_j}p_{c_j}}{v^*\sigma_c^2(\mathbf{P}_r)+N_cl_{c_j}p_{c_j}}.
\end{equation}
The aforementioned approximation is derived under the assumption that transmitters and receivers are equipped with an infinite number of antennas. In this limit, the random channel matrix tends to be determined. Leveraging the random matrix theory, we can extract the expectation operation and derive this approximation. Readers can refer to \cite{reff} for a detailed proof. On this basis, we transform (P1) into the following problem,
\begin{subequations}
	\label{P2-A}
	\begin{align}
		\mbox{(P2)} \max\limits_{\mathbf{P}_c,\mathbf{P}_r }&\min\limits_{i\in[1,M_r]}\
		\rho_i(\mathbf{P}_r, \mathbf{P}_c)  \\
		s.t. \ \ &\bar{R}_{ap}(\mathbf{P}_c, \mathbf{P}_r,v^*) \geqslant R_{req} \label{r1}\\
		%\ \ \ \ \ \ &\rho_i(\mathbf{P}_r, \mathbf{P}_c) \geqslant \gamma, i=1,...,M_r \\
		\ \ \ \ \ \ &\frac{1}{v^*}+\sum\limits_{j=1}^{M_c}\frac{l_{c_j}p_{c_j}}{v^*\sigma_c^2(\mathbf{P}_r)+N_cl_{c_j}p_{c_j}}=1 \label{r2}\\
		&\mathbf{0}_{M_c} \preceq \mathbf{P}_c  \preceq P_{cmax}\cdot\mathbf{I}_{M_c}\\
		&\mathbf{0}_{M_r} \preceq \mathbf{P}_r  \preceq P_{rmax}\cdot\mathbf{I}_{M_r} \\
		&\operatorname{tr}(\mathbf{P}_c)\leqslant P_{csum} \\
		&\operatorname{tr}(\mathbf{P}_r)\leqslant P_{rsum}.
		%\ \ \ \ \ \ &\eqref{4}-\eqref{5}. \nonumber
	\end{align}
\end{subequations}
%where $\gamma$ is an introduced slack variable and \eqref{r2} is just \eqref{10}.
However, $\bar{R}_{ap}(\mathbf{P}_c,\mathbf{P}_r,v^*)$ remains a complex expression. It includes a fixed point $v^*$, which is determined by the nonlinear equality \eqref{r2}. To address this challenge, we apply auxiliary functions and scaling techniques, as detailed in the next subsection.
\subsubsection{AUXILIARY-FUNCTION-BASED SCALING}
We use the right side of \eqref{10} to replace the term $[1-\frac{1}{v^*}]$ in \eqref{11}. This introduces an auxiliary function, denoted as  $g(\mathbf{P}_c, \mathbf{P}_r, z)$:
\begin{equation}
	\label{12}
	\begin{aligned}
		g(\mathbf{P}_c, \mathbf{P}_r, z)=&\sum\limits_{j=1}^{M_c}\log_2\big(1+\frac{N_cl_{c_j}p_{c_j}}{z\sigma_c^2(\mathbf{P}_r)}\big)+N_c\log_2(z)\\
		-&\sum\limits_{j=1}^{M_c}N_c\log_2(e)\big[\frac{l_{c_j}p_{c_j}}{z\sigma_c^2(\mathbf{P}_r)+N_cl_{c_j}p_{c_j}}\big].
	\end{aligned}
\end{equation}
Obviously, when $z=v^*$, $g(\mathbf{P}_c, \mathbf{P}_r, z)$ has the following connection with the ergodic rate expression:
\begin{equation}
	\label{r6}
	\bar{R}_{ap}(\mathbf{P}_c,\mathbf{P}_r,v^*)=g(\mathbf{P}_c, \mathbf{P}_r, z)\bigg|_{z=v^*}.
\end{equation}
In addition, we propose the following \textbf{Lemma}.
\begin{lemma} On the condition of $N_c\geqslant M_c$, the following equation holds:
	\begin{equation}
		\label{2}
		\bar{R}_{ap}(\mathbf{P}_c,\mathbf{P}_r,v^*)=\max_{1\leqslant z\leqslant v^*}g(\mathbf{P}_c, \mathbf{P}_r, z).
	\end{equation}
	%$g(\mathbf{P}_c, \mathbf{P}_r, z)$ is a monotone increasing function of $z$ on the condition of  $N_c \geqslant M_c$.
\end{lemma}
\begin{Proof}
	See Appendix A.
\end{Proof}
%Based on the increasing monotonicity property of $g(\mathbf{P}_c, \mathbf{P}_r, z)$ in terms of $z$, we have that
%\begin{equation}
%	\bar{R}_{ap}(\mathbf{P}_c,\mathbf{P}_r,v^*)=\max\limits_{1\leqslant z \leqslant v^*}g(\mathbf{P}_c, \mathbf{P}_r, z).
%\end{equation}
Based on the lemma, the ergodic rate constraint \eqref{r1} can be converted into a less-than-max inequality and further scaled into
\begin{equation}
	%\nonumber
	\label{eq111}
	\begin{aligned}
		\bar{R}_{ap}(\mathbf{P}_c,\mathbf{P}_r,v^*)\geqslant R_{req} &\Rightarrow\max\limits_{1\leqslant z \leqslant v^*}g(\mathbf{P}_c, \mathbf{P}_r, z)\geqslant R_{req} \\
		&\Rightarrow
		\left\{
		\begin{split}
			&g(\mathbf{P}_c, \mathbf{P}_r, z) \geqslant R_{req},\\
			&1\leqslant z \leqslant v^*.
		\end{split}
		\right.
	\end{aligned}
\end{equation}
On this basis, we rewrite (P2) as
\begin{subequations}
	\label{P3}
	\begin{align}
		\mbox{(P3)}  \max\limits_{\mathbf{P}_c,\mathbf{P}_r, z }&\min\limits_{i\in[1,M_r]}\
		\rho_i(\mathbf{P}_r, \mathbf{P}_c)  \\
		s.t. \ \ &g(\mathbf{P}_c, \mathbf{P}_r, z) \geqslant R_{req}\\
		%	\ \ \ \ \ \ &\rho_i(\mathbf{P}_r, \mathbf{P}_c) \geqslant \gamma, i=1,...,M_r \\
		\ \ \ \ \ \
		&1\leqslant z \leqslant v^*\\
		&\frac{1}{v^*}+\sum\limits_{j=1}^{M_c}\frac{l_{c_j}p_{c_j}}{v^*\sigma_c^2(\mathbf{P}_r)+N_cl_{c_j}p_{c_j}}=1 \label{r5}\\
		&\mathbf{0}_{M_c} \preceq \mathbf{P}_c  \preceq 	P_{cmax}\cdot\mathbf{I}_{M_c} \\
		&\mathbf{0}_{M_r} \preceq \mathbf{P}_r  \preceq P_{rmax}\cdot\mathbf{I}_{M_r} \\
		&\operatorname{tr}(\mathbf{P}_c)\leqslant P_{csum} \\
		&\operatorname{tr}(\mathbf{P}_r)\leqslant P_{rsum}.
	\end{align}
\end{subequations}
Moving forward, we reorganize $g(\mathbf{P}_c, \mathbf{P}_r, z)$ as
\begin{equation}
	\begin{aligned}
		g(\mathbf{P}_c, \mathbf{P}_r, z)=
		-\log_2(e)\sum\limits_{j=1}^{M_c}\bigg(\frac{N_cl_{c_j}p_{c_j}}{z\sigma_c^2(\mathbf{P}_r)+N_cl_{c_j}p_{c_j}}+&\\
		\mbox{In}\big(1-\frac{N_cl_{c_j}p_{c_j}}{z\sigma_c^2(\mathbf{P}_r)+N_cl_{c_j}p_{c_j}}\big)\bigg)+N_c\log_2z.&
	\end{aligned}
\end{equation}
Through observations, it is easy to find that the fraction term, i.e., $\frac{N_cl_{c_j}p_{c_j}}{z\sigma_c^2(\mathbf{P}_r)+N_cl_{c_j}p_{c_j}}$, occurs twice in $g(\mathbf{P}_c, \mathbf{P}_r, z)$. Motivated by this, we  introduce a group of variables, denoted as  $\mathbf{t}=\{t_j\}_{j=1}^{M_c}$, to replace these fraction terms. Then, we transfer the auxiliary function $g(\mathbf{P}_c, \mathbf{P}_r, z)$ into $\mathcal{G}(z, \mathbf{t})$:
\begin{equation}
	\begin{aligned}
		\mathcal{G}(z, \mathbf{t})=
		-\log_2(e)\sum\limits_{j=1}^{M_c}\big(t_j+
		\mbox{In}(1-t_j)\big)+N_c\log_2z.
	\end{aligned}
\end{equation}
Correspondingly, we denote $\mathbf{t}^*=\{t_j^*\}_{j=1}^{M_c}$ and $t_j^*$ is given by
\begin{equation}
	\label{r7}
	t^*_j\triangleq\frac{N_cl_{c_j}p_{c_j}}{z\sigma_c^2(\mathbf{P}_r)+N_cl_{c_j}p_{c_j}}.
\end{equation}
Similar to \eqref{r6},  when $\mathbf{t}=\mathbf{t}^*$ and $z=v^*$, we have
\begin{equation}
	\bar{R}_{ap}(\mathbf{P}_c,\mathbf{P}_r,v^*)=g(\mathbf{P}_c, \mathbf{P}_r, z)\bigg|_{z=v^*}=\mathcal{G}(z, \mathbf{t})\bigg|_{z=v^*, \mathbf{t}=\mathbf{t}^*}.
\end{equation}
%\end{subequations}
Since  $-[x+\text{In}(1-x)]$ and $\log(x)$ are monotonically increasing with $x$, it is easy to find that $\mathcal{G}(z, \mathbf{t})$ is an increasing function of $\mathbf{t}$ and $z$. Therefore, we have the following equation,
\begin{equation}
	g(\mathbf{P}_c, \mathbf{P}_r, z)=\max_{0\leqslant t_j\leqslant t^*_j,\forall j} \mathcal{G}(z, \mathbf{t}).
\end{equation}
On this basis, the ergodic rate constraint can be further scaled into
\begin{equation}
	\label{eq122}
	\begin{aligned}
		\bar{R}_{ap}(\mathbf{P}_c,\mathbf{P}_r,v^*)\geqslant R_{req} &\Rightarrow\max\limits_{1\leqslant z \leqslant v^*}g(\mathbf{P}_c, \mathbf{P}_r, z)\geqslant R_{req} \\
		&\Rightarrow\max\limits_{1\leqslant z\leqslant v^*, 0\leqslant t_j\leqslant t^*_j,\forall j}\mathcal{G}(z, \mathbf{t})\geqslant R_{req}\\
		&\Rightarrow
		\left\{
		\begin{split}
			&\mathcal{G}(z, \mathbf{t}) \geqslant R_{req},\\
			&1\leqslant z \leqslant v^*,\\
			&0\leqslant t_j\leqslant t^*_j, j=1,...,M_c.
		\end{split}
		\right.
	\end{aligned}
\end{equation}
Based on \eqref{eq122}, we rewrite (P3) into (P4)
\begin{subequations}
	\label{P4}
	\begin{align}
		\mbox{(P4)}  \max\limits_{\mathbf{P}_c,\mathbf{P}_r,z, \mathbf{t} }&\min\limits_{i\in[1,M_r]}\
		\rho_i(\mathbf{P}_r, \mathbf{P}_c)  \\
		s.t. \ \ &\mathcal{G}(z,\mathbf{t}) \geqslant R_{req} \label{r3}\\
		&1\leqslant z \leqslant v^*\label{r4}\\
		&0\leqslant t_j\leqslant t^*_j, j=1,...,M_c \\
		&\frac{1}{v^*}+\sum\limits_{j=1}^{M_c}\frac{l_{c_j}p_{c_j}}{v^*\sigma_c^2(\mathbf{P}_r)+N_cl_{c_j}p_{c_j}}=1\\
		&\mathbf{0}_{M_c} \preceq \mathbf{P}_c  \preceq 	P_{cmax}\cdot\mathbf{I}_{M_c}\\
		&\mathbf{0}_{M_r} \preceq \mathbf{P}_r  \preceq P_{rmax}\cdot\mathbf{I}_{M_r} \\
		&\operatorname{tr}(\mathbf{P}_c)\leqslant P_{csum} \\
		&\operatorname{tr}(\mathbf{P}_r)\leqslant P_{rsum}.
	\end{align}
\end{subequations}
To further simplify this problem, we rewrite the fixed-point equation \eqref{10} as follows:
\begin{equation} \frac{1}{v^*}+\sum\limits_{j=1}^{M_c}\frac{l_{c_j}p_{c_j}}{v^*\sigma_c^2(\mathbf{P}_r)+N_cl_{c_j}p_{c_j}} = 1.
\end{equation}
Since $\frac{1}{z}+\sum\limits_{j=1}^{M_c}\frac{l_{c_j}p_{c_j}}{z\sigma_c^2(\mathbf{P}_r)+N_cl_{c_j}p_{c_j}}$ is monotonically decreasing with $z$,  $z \leqslant v^*$ is equivalent to
\begin{equation}
	\begin{aligned}
		\frac{1}{z}+\sum\limits_{j=1}^{M_c}\frac{l_{c_j}p_{c_j}}{z\sigma_c^2(\mathbf{P}_r)+N_cl_{c_j}p_{c_j}} \geqslant 1,
	\end{aligned}
\end{equation}
which can be further simplified as
\begin{align*}
	\frac{1}{z}+\sum\limits_{j=1}^{M_c}&\frac{l_{c_j}p_{c_j}}{z\sigma_c^2(\mathbf{P}_r)+N_cl_{c_j}p_{c_j}} \geqslant 1\Rightarrow \frac{1}{z}+\frac{1}{N_c}\sum\limits_{j=1}^{M_c}t^*_j\geqslant 1\\ &\Rightarrow
	\left\{
	\begin{aligned}
		&\frac{1}{z}+\frac{1}{N_c}\sum\limits_{j=1}^{M_c}t_j\geqslant 1,\\
		&t_j\leqslant t_j^*, j=1,...,M_c.
	\end{aligned}\right.
\end{align*}
For $t_j\leqslant t_j^*$, the logarithmic operation is applied to simplify this constraint,
\begin{equation}
	\begin{aligned}
		t_j\leqslant t_j^*&\Rightarrow t_j\leqslant \frac{N_cl_{c_j}p_{c_j}}{z\sigma_c^2(\mathbf{P}_r)+N_cl_{c_j}p_{c_j}}\\
		&\Rightarrow t_j[z\sigma_c^2(\mathbf{P}_r)+N_cl_{c_j}p_{c_j}] \leqslant N_cl_{c_j}p_{c_j} \\
		&\Rightarrow t_jz\sigma_c^2(\mathbf{P}_r) \leqslant (1-t_j)N_cl_{c_j}p_{c_j}\\
		&\Rightarrow
		\mbox{In}(t_j)+\mbox{In}(z)+\mbox{In}(\sigma_c^2(\mathbf{P}_r))\\
		&\quad\quad\leqslant\mbox{In}(1-t_j)+\mbox{In}(N_cl_{c_j}p_{c_j}). \label{r8}
	\end{aligned}
\end{equation}
On this basis, we transform (P4) into (P5)
\begin{subequations}
	\label{P5}
	\begin{align}
		\mbox{(P5)} \max\limits_{\mathbf{P}_c,\mathbf{P}_r,z, \mathbf{t} }&\min\limits_{i\in[1,M_r]} \rho_i(\mathbf{P}_c,\mathbf{P}_r)\\
		s.t. \ \ &\mathcal{G}(z, \mathbf{t}) \geqslant R_{req} \label{r11}\\
		&\frac{1}{z}+\frac{1}{N_c}\sum\limits_{j=1}^{M_c}t_j\geqslant 1\label{r12}\\
		&\mbox{In}(1-t_j)+\mbox{In}(N_cl_{c_j}p_{c_j}) \geqslant \label{r13} \\
		&\ \ \ \ \ \ \quad \mbox{In}(t_j)+\mbox{In}(z)+\mbox{In}(\sigma_c^2(\mathbf{P}_r)) \nonumber\\
		&z\geqslant1\\
		&t_j\geqslant0, j=1,...,M_c\\
		&\mathbf{0}_{M_c} \preceq \mathbf{P}_c  \preceq 	P_{cmax}\cdot\mathbf{I}_{M_c}\\
		&\mathbf{0}_{M_r} \preceq \mathbf{P}_r  \preceq P_{rmax}\cdot\mathbf{I}_{M_r} \\
		&\operatorname{tr}(\mathbf{P}_c)\leqslant P_{csum} \\
		&\operatorname{tr}(\mathbf{P}_r)\leqslant P_{rsum}.
	\end{align}
\end{subequations}
Moving forward, we address the non-convex objective. This max-min optimization can be equivalently rewritten as
\begin{subequations}
	\begin{align}
		\max\limits_{\mathbf{P}_c,\mathbf{P}_r,z, \mathbf{t} } &\gamma\\
		\mbox{s.t.} \ &\rho_i(\mathbf{P}_c,\mathbf{P}_r) \geqslant \gamma, i=1,....,M_r,  \label{r9}
	\end{align}
\end{subequations}
where $\gamma$ is an introduced auxiliary variable. Since $\rho_i(\mathbf{P}_r, \mathbf{P}_c)$ is a ratio expression of $\mathbf{P}_c$ and $\mathbf{P}_r$,  \eqref{r9} can be tackled by the fractional programming scheme named quadratic transform \cite{ref16},
\begin{equation}
	\label{41}
	\max\limits_{\beta_i} \bigg[2\beta_i\sqrt{[h_{{r_i}\rightarrow {r_i}}]^Hp_{r_i}h_{{r_i}\rightarrow {r_i}}}-\beta_i^2\sigma^2_{r_i}(\mathbf{P}_c)\bigg]\geqslant\ \gamma,
\end{equation}
where $\beta_i$ is an auxiliary variable. The solution to \eqref{41}, denoted as $\beta_i^*$, can be calculated by a closed-form expression,
\begin{equation}
	\beta_i^*=\frac{\sqrt{[h_{{r_i}\rightarrow {r_i}}]^Hp_{r_i}h_{{r_i}\rightarrow {r_i}}}}{\sigma^2_{r_i}(\mathbf{P}_c)}, i=1,...,M_r. \label{21}
\end{equation}

\begin{algorithm}[t]
	\caption{Iterative Algorithm for Large-Scale-CSI-Based C\&S Power Allocation.}
	\label{tab1}
	\begin{algorithmic}[1]
		\REQUIRE {$P_{cmax}$, $P_{rmax}$, $P_{csum}$, $P_{rsum}$, $R_{req}$, $M_c$, $M_r$, $N_c$, and the termination threshold for iterations, $\epsilon$}.
		\STATE \emph{Initialization}: $(\mathbf{P}_c)^0=\min\{\frac{P_{csum}}{M_c}, P_{cmax}\}\mathbf{I}_{M_c}$, $(\mathbf{P}_{r})^0=\mathbf{0}_{M_r}$, $(\gamma)^0=0$, and $s=0$.
		\STATE Calculate $v^*$ and $\mathbf{t}^*$ according to \eqref{10} and \eqref{r7}, and let $(z)^0=v^*$ and $(\mathbf{t})^0=\mathbf{t}^*$.
		\REPEAT
		\STATE  Calculate  $(\beta_i^*)^s\ (i=1,...,M_r)$ according to \eqref{21};
		\STATE  $s=s+1$;
		\STATE  Solve (P6) and update  $(\mathbf{P}_c)^s$, $(\mathbf{P}_r)^s$,  $(z)^s$, $(\mathbf{t})^s$, and $(\gamma)^s$;
		\UNTIL{$\frac{\left|(\gamma)^s-(\gamma)^{s-1} \right|}{(\gamma)^s}  \ \leqslant \epsilon$.}
		\ENSURE  $(\mathbf{P}_c)^s$ and $(\mathbf{P}_{r})^s$.
	\end{algorithmic}
\end{algorithm}

\begin{figure*}
	\begin{equation}
		\mathcal{G}(z,\mathbf{t}|(\mathbf{t})^{s-1})=-\log_2(e)\sum_{j=1}^{M_c}\bigg(\frac{(t_j)^{s-1}(1-t_j)}{1-(t_j)^{s-1}}+\mbox{In}(1-(t_j)^{s-1})\bigg)+N_c\log_2z
	\end{equation}
	%\begin{equation}
	%	\rho_i(\mathbf{P}_r, \mathbf{P}_c|\mathbf{P}_r^{s-1},\mathbf{P}_c^{s-1} )=\frac{l_{ii}^rp_{r_i}}{\sigma_{r_i}^2(\mathbf{P}_c^{s-1})}-\frac{l_{ii}^r(p_{r_i})^{s-1}\sum\limits_{j=1}^{M_c}l_{ij}^{cr}\big(p_{c_j}-(p_{c_j})^{s-1}\big)}{\sigma_{r_i}^4(\mathbf{P}_c^{s-1})}
	%\end{equation}
	\begin{equation}
		\begin{aligned}
			&\mbox{In}\big(t_j|(t_j)^{s-1}\big)+\mbox{In}\big(z|(z)^{s-1}\big)  +\mbox{In}\big(\sigma_c^2(\mathbf{P}_r)|\sigma_c^2((\mathbf{P}_r)^{s-1})\big)=\\
			&\quad\quad\quad\quad\frac{t_j-(t_j)^{s-1}}{(t_j)^{s-1}}+\frac{z-(z)^{s-1}}{(z)^{s-1}}+\frac{\sum\limits_{i=1}^{M_r}l_{r_i\rightarrow c}(p_{r_i}-(p_{r_i})^{s-1})}{\sigma_c^2((\mathbf{P}_r)^{s-1})}+\mbox{In}\big((t_j)^{s-1}(z)^{s-1}\sigma_c^2((\mathbf{P}_r)^{s-1})\big)
		\end{aligned}
	\end{equation}
	\hrulefill
\end{figure*}

On this basis, we propose our iterative algorithm. Taking iteration $s$ as an example, we calculate $(\beta_i^*)^{s-1}$ according to \eqref{21} using the transmit power obtained in iteration $s-1$, i.e., $(\mathbf{P_c}^*)^{s-1}$ and $(\mathbf{P_r}^*)^{s-1}$.  Then, we use it to replace $\beta_i$ in the iteration $s$ and \eqref{41} is simplified as
\begin{equation}
	2(\beta_i^*)^{s-1}\sqrt{[h_{{r_i}\rightarrow {r_i}}]^Hp_{r_i}h_{{r_i}\rightarrow {r_i}}}-((\beta_i^*)^{s-1})^2\sigma^2_{r_i}(\mathbf{P}_c)\geqslant\ \gamma.
\end{equation}
As for  constraints \eqref{r11}, \eqref{r12} and \eqref{r13}, the Taylor expansion is used to linearize the non-convex terms therein. On this basis, (P5) is transformed into a series of iterative convex problems. The problem in iteration $s$ is given by
\begin{subequations}
	\begin{align}
		\mbox{(P6)} \  &\max\limits_{\mathbf{P}_c,\mathbf{P}_r, z, \mathbf{t}, \gamma} \gamma \label{eq113}\\
		s.t. \ \ &\mathcal{G}(z,\mathbf{t}|\mathbf{t}^{s-1}) \geqslant R_{req}\\
		&2(\beta_i^*)^{s-1}\sqrt{[h_{{r_i}\rightarrow {r_i}}]^Hp_{r_i}h_{{r_i}\rightarrow {r_i}}}\nonumber\\
		&\quad\quad\quad-((\beta_i^*)^{s-1})^2\sigma^2_{r_i}(\mathbf{P}_c)\geqslant\ \gamma \label{eq112}\\
		&\frac{2}{z^{s-1}}-\frac{z}{(z^{s-1})^2}+\frac{1}{M_c}\sum\limits_{j=1}^{M_c}t_j\geqslant 1 \\
		&\mbox{In}(N_cl_{c_j}p_{c_j})+\mbox{In}(1-t_j)\geqslant \mbox{In}\big(t_j|t_j^{s-1}\big) \\
		&\quad\quad\quad\quad+\mbox{In}\big(z|z^{s-1}\big)  +\mbox{In}\big(\sigma_c^2(\mathbf{P}_r)|\sigma_c^2(\mathbf{P}_r^{s-1})\big)\nonumber\\
		&z \geqslant 1\\
		& t_j \geqslant 0, \ j=1,...,M_c\\
		&\mathbf{0}_{M_c} \preceq \mathbf{P}_c  \preceq 	P_{cmax}\cdot\mathbf{I}_{M_c} \\
		&\mathbf{0}_{M_r} \preceq \mathbf{P}_r  \preceq P_{rmax}\cdot\mathbf{I}_{M_r} \\
		&\operatorname{tr}(\mathbf{P}_c)\leqslant P_{csum} \\
		&\operatorname{tr}(\mathbf{P}_r)\leqslant P_{rsum},
	\end{align}
\end{subequations}
where $\mathcal{G}(z,\mathbf{t}|\mathbf{t}^{s-1})$ and $\mbox{In}\big(t_j|t_j^{s-1}\big)+\mbox{In}\big(z|z^{s-1})\big)  +\mbox{In}\big(\sigma_c^2(\mathbf{P}_r)|\sigma_c^2((\mathbf{P}_r)^{s-1})\big)$ are the Taylor expansion expressions of $\mathcal{G}(z,\mathbf{t})$ and $\mbox{In}(t_j)+\mbox{In}(z)+\mbox{In}(\sigma_c^2(\mathbf{P}_r))$. Their expressions are given at the top of the this page. The detailed algorithm  is summarized in \textbf{Algorithm \ref{tab1}}. The proof of convergence is given in Appendix B. In addition, we analyze the computation complexity of the proposed algorithm. (P6) is a convex problem.  If the classical interior point method is employed to solve it, the computational complexity is given by $\mathcal{O}\left(({2M_c+M_r+2})^{3.5}\log(1/\epsilon)\right)$, where $(2M_c+M_r+2)$ represents the dimension of the optimization variables and $\epsilon_0$ is the target accuracy. Assuming that the proposed algorithm converges within $I$ iterations, the total computational complexity is thus $\mathcal{O}\left(I({2M_c+M_r+2})^{3.5}\log(1/\epsilon)\right)$.

\begin{figure}[t]
	\centering
	\includegraphics[width=0.42\textwidth]{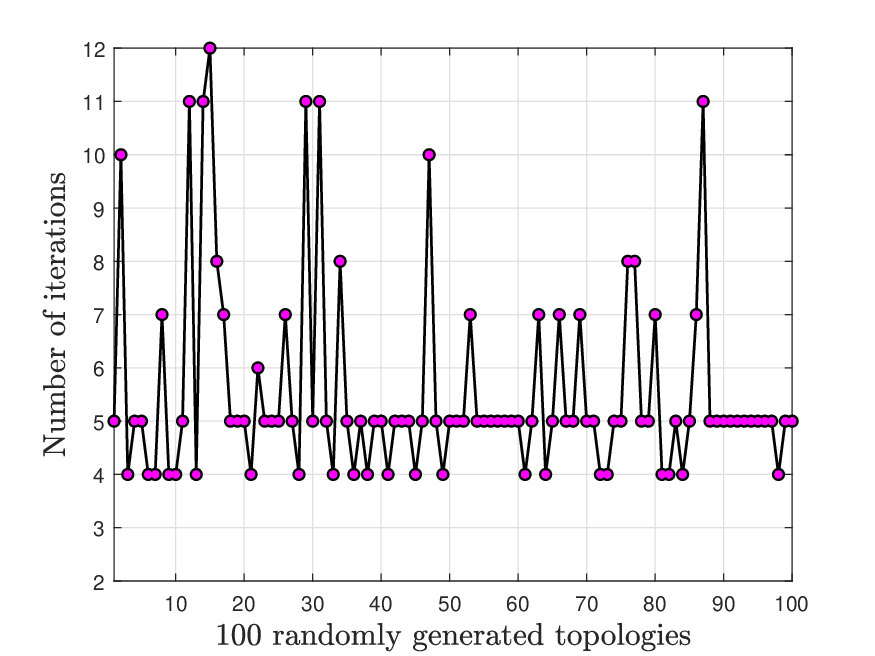}
	\caption{The number of iterations of the proposed iterative algorithm under 100 randomly generated typologies.}
	\label{fig6}
\end{figure}

	\section{SIMULATION RESULTS AND DISCUSSIONS}
	\label{sec4}

	\begin{figure}[t]
		\centering
		\includegraphics[width=0.42\textwidth]{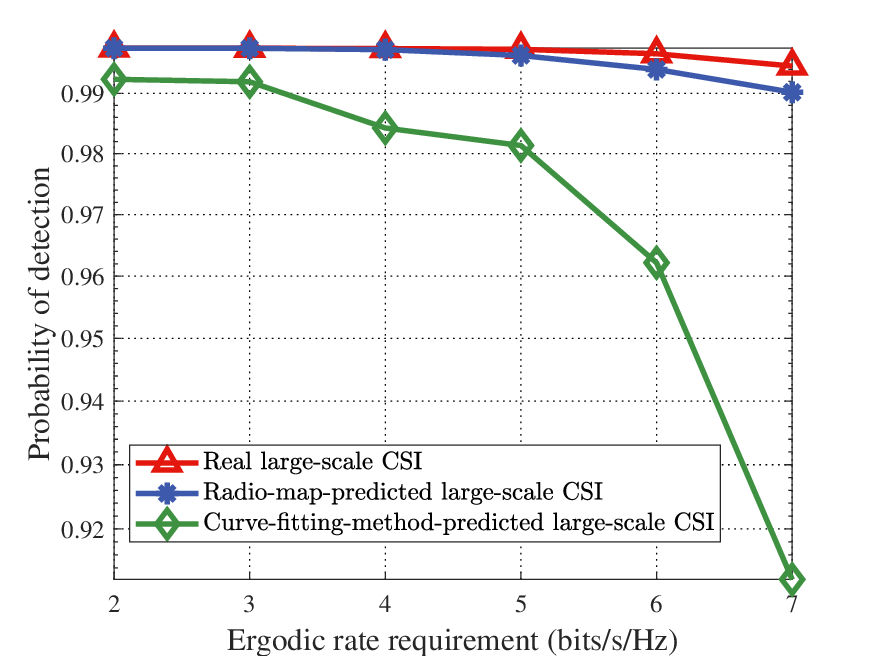}
		\caption{The radar detection performance under different CSI data, varies with the ergodic rate requirement.}
		\label{fig1-1}
	\end{figure}

	In this section, we present  simulation results and discussions. The simulation layout is the same as that we established in the case study of the radio map,  as presented in Fig. \ref{fig1} (a).  The layout includes three BSs, i.e., $M_c=3$,  two radars, i.e., $M_r=2$, one interfered user, and a red diamond-shaped target.  They are located at (604 \text{m}, 629 \text{m}), (1289 \text{m}, 2022 \text{m}), (1986 \text{m}, 1316 \text{m}), (-1167 \text{m}, 3125 \text{m}), (2620 \text{m}, -779 \text{m}), (650 \text{m}, 1134 \text{m}), and (1360 \text{m}, 1000 \text{m}), respectively.  BSs and the interfered user are equipped with half-wave dipole antennas. The antenna number of the user is equal to the number of BSs, i.e, $N_c=3$. Radars are equipped with directional antennas. The maximum antenna gain is $30$ dBi and the half-power beam width, denoted as $\theta_{\text{3dB}}$, is $32^{\circ}$. The carrier frequency is $f_c=2.8$ GHz, which is the frequency band of the surveillance radar, and the noise variance is $\sigma^2=-107 \ \text{dBm}$ \cite{add}. Power-related parameters are set as $P_{cmax}=40$ W, $P_{csum}=100$ W, $P_{rmax}=1$ kW and $P_{rsum}=1.5$ kW. This is because radars have much higher power than BSs \cite{ref12}. The sampling number of radars is $N=256$ per CPI, the false alarm probability is $P_{F}=1\times10^{-4}$, and thus, $\mu=0.0182$ according to \eqref{eq1}.

	We first present the convergence performance of the proposed iterative algorithm. In this simulation, we randomly changed the user and target positions and ran the proposed algorithm. The number of iterations was recorded and presented in Fig. \ref{fig6}.
	Remarkably, in over $90\%$ cases, the proposed algorithm converges within 10 times. The maximum number of iterations observed is only $12$. These results affirm the rapid convergence of the proposed algorithm.

	\begin{figure}[t]
		\centering
		\includegraphics[width=0.5\textwidth]{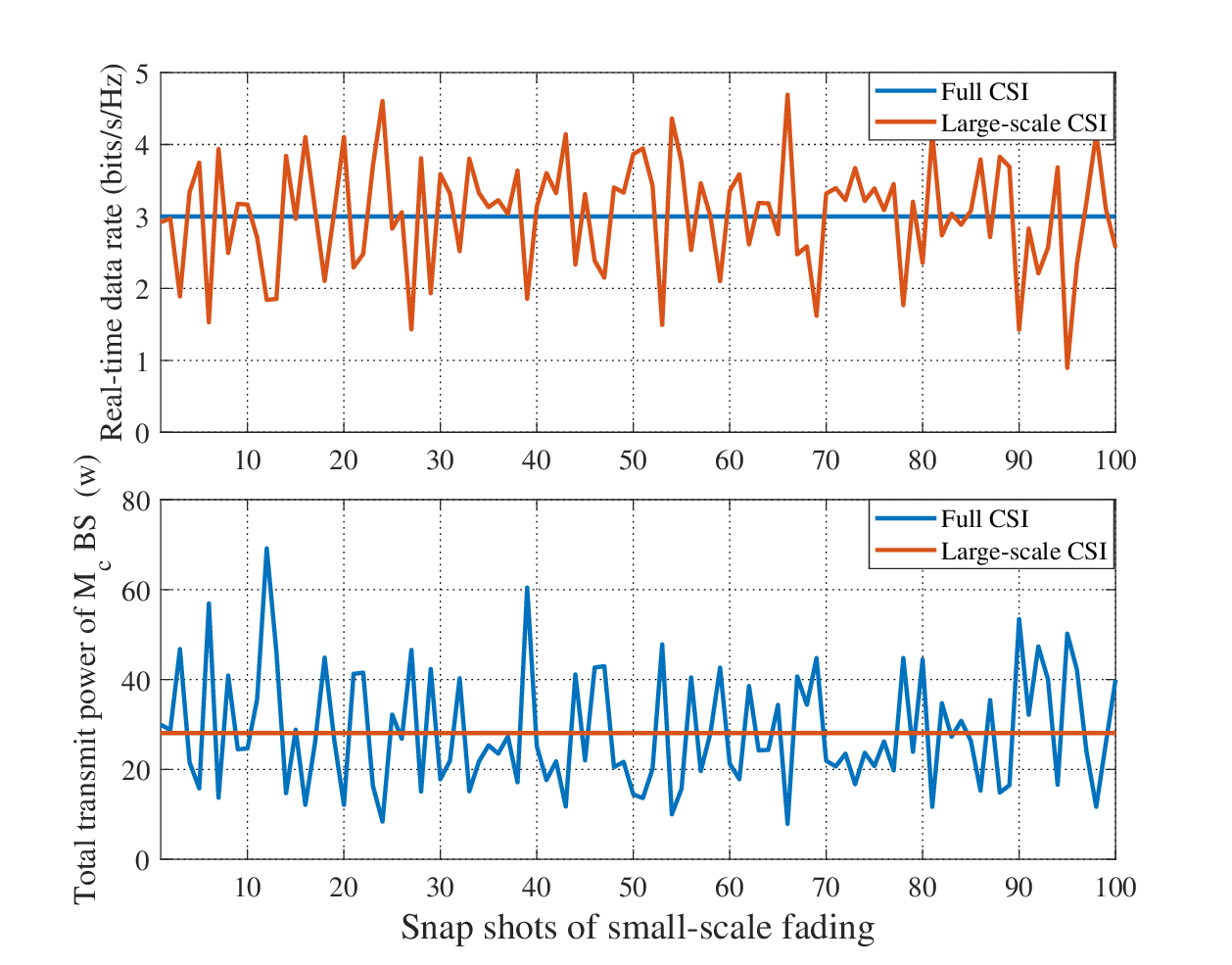}
		\caption{Comparisons between the large-scale-CSI-based scheme and the full-CSI-based scheme}
		\label{figr2}
	\end{figure}

	\begin{figure}[t]
		\centering
		\includegraphics[width=0.48\textwidth]{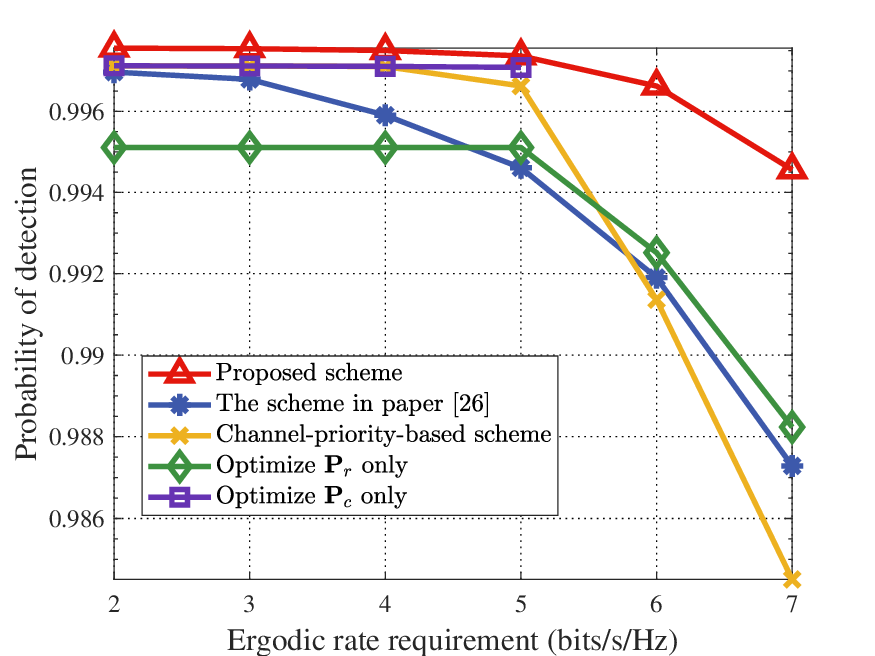}
		\caption{The radar detection performance under different schemes, varies with the ergodic rate requirement.}
		\label{fig2}
	\end{figure}

	We evaluate the impact of large-scale CSI accuracy on C\&S performance in Fig. \ref{fig1-1}. It can be seen that the curve corresponding to the radio-map-predicted large-scale CSI closely aligns with the one using the real data, while the curve obtained through the curve-fitting method exhibits a noticeable gap. This discrepancy is expected, given that the radio map provides a more accurate estimation of the real large-scale CSI, as demonstrated in Fig. \ref{fig1}. In our case study, the average deviation of the large-scale CSI map generated by the radio map is below $4$ dB while the deviation associated with the curve-fitting method exceeds $25$ dB. Such a substantial deviation leads to misinformed power allocation and, consequently, performance deterioration.  Therefore, compared with the channel model, which is blind to the surrounding environment, the radio map verifies  the extrinsic information, i.e., positions and surroundings, is effective in coordinating conflicting signals.

	In Fig. \ref{figr2}, we show the difference of using large-scale CSI and full CSI. In this simulation, we set the rate requirement of the interfered user to $3$ bits/s/Hz and assume that the varying frequency of the small-scale fading is one hundred times that of the large-scale fading.  The small-scale fading is randomly generated following the Gaussian distribution, i.e., $\mathcal{CN}(0,1)$. Fig. \ref{figr2} presents one snapshot of the large-scale CSI, which corresponds to $100$ snapshots of the small-scale CSI. From the top subfigure, we can see that, the large-scale-CSI-based scheme cannot guarantee the real-time rate as effectively as the full-CSI-based scheme. This is the compromise of using the large-scale CSI. In return, from the bottom subfigure we can see that the large-scale-CSI-based scheme makes adjustments in large-scale intervals, while the full-CSI-based scheme adjusts the systems in small-scale intervals.  This demonstrates that our large-scale-CSI-based scheme significantly lowers the negotiation and adjustment frequency of C\&S systems, allowing C\&S systems to operate in a loosely cooperated manner.

\begin{algorithm}[t]
	\caption{C\&S Power Allocation Algorithm According to \cite{ref11} and  Channel-Priority-Based Scheme.}
	\label{tab2}
	\begin{algorithmic}[1]
		\REQUIRE {$P_{cmax}$, $P_{rmax}$, $P_{csum}$, $P_{rsum}$, $R_{req}$, $M_c$, $M_r$, $N_c$, and                                                                                                                                                                                                                                                                                                                                                                                                                                                                                                                                                      step parameter $\Delta_P$. }
		\STATE $\mathbf{P}_c=\min\{\frac{P_{csum}}{M_c}, P_{cmax}\}\mathbf{I}_{M_c}$, $\mathbf{P}_{r}=\min\{\frac{P_{rsum}}{M_r}, P_{rmax}\}\mathbf{I}_{M_r}$.
		\REPEAT
		\STATE Calculate $v^*$ and $\bar{R}_{ap}(\mathbf{P}_c,\mathbf{P}_r,v^*)$ according to \eqref{10} and \eqref{11}.
		\IF{$\bar{R}_{ap}(\mathbf{P}_c,\mathbf{P}_r,v^*)<R_{req}$}
		\STATE  The scheme proposed in \cite{ref11}:\\ {\setlength{\parindent}{2em}$p_{r_i}=\max\{p_{r_i}-\frac{1}{M_r}\Delta_P,0\},  \forall i$} \\
		The channel-priority-based scheme:\\ {\setlength{\parindent}{2em}$p_{r_i}=\max\{p_{r_i}-\frac{l_{r_i\rightarrow c}p_{r_i}}{\sum\limits_{i=1}^{M_r}l_{r_i\rightarrow c}p_{r_i}}\Delta_P,0\}, \forall i$}
		\ELSIF{$\bar{R}_{ap}(\mathbf{P}_c,\mathbf{P}_r,v^*)>R_{req}$}
		\STATE  The scheme proposed in \cite{ref11}:\\ {\setlength{\parindent}{2em}$p_{c_i}=\max\{p_{c_i}-\frac{1}{M_c}\Delta_P,0\}, \forall i$} \\
		The channel-priority-based scheme:\\ {\setlength{\parindent}{2em}$p_{c_i}=\max\{p_{c_i}-\frac{\sum\limits_{j=1}^{M_r}l_{c_i\rightarrow r_j}p_{c_i}}{\sum\limits_{i=1}^{M_c}\sum\limits_{j=1}^{M_r}l_{c_i\rightarrow r_j}p_{c_i}}\Delta_P,0\},\forall i$}
		\ENDIF
		\UNTIL{$\bar{R}_{ap}(\mathbf{P}_c,\mathbf{P}_r,v^*)=R_{req}$}
		\ENSURE  $\mathbf{P}_c$ and $\mathbf{P}_{r}$.
	\end{algorithmic}

\end{algorithm}

	In Fig. \ref{fig2}, we compare the proposed scheme with the scheme proposed in \cite{ref11}, the channel-priority-based scheme, and the unilateral schemes of optimizing $\mathbf{P}_c$ or $\mathbf{P}_r$ only. The scheme proposed in \cite{ref11} considered the power allocation between one BS and one radar.
	Since we consider a multi-BS and multi-radar network, the equal power allocation is used among BSs and radars as the initial allocation, i.e., $\mathbf{P}_c=\min\{\frac{P_{csum}}{M_c}, P_{cmax}\}\mathbf{I}_{M_c}$ and $\mathbf{P}_r=\min\{\frac{P_{rsum}}{M_r},P_{rmax}\}\mathbf{I}_{M_r}$, and the step parameter is set as $\Delta_p=0.5$ w. Then, per the conclusion drawn in \cite{ref11}, that the optimal power allocation must be that one of C\&S systems uses up its available power and the other side reserves power, we thus test the ergodic-rate condition to decide which side should reserve its power.  As for the channel-priority-based scheme, we also use equal power allocation as its initial allocation and leverage large-scale CSI of the interference channels to adjust C\&S power. Their detailed algorithms are summarized in {\bf{Algorithm 2}}. From Fig. \ref{fig2}, we can see that the proposed scheme achieves the highest detection probability. This is because the proposed scheme leverages the large-scale CSI and intelligently allocates more power to the BSs and radars with favorable serving channels and unfavorable interference channels. This amplifies the positive effects of power for their own systems while constraining the negative effects on the opposite system. In contrast, the scheme proposed in \cite{ref11} is blind to the large-scale CSI, treating all BSs and radars equally in the power allocation. This uniform allocation cannot avoid signal collisions, especially when the transmitting equipment on one side and the receiving equipment on the other side are in close proximity. Only in the special case of one BS and one radar, the scheme proposed in \cite{ref11} gives the optimal solution. As for the channel-priority-based scheme, when $R_{req}\leqslant 5 $ bits/s/Hz, its performance is close to that of the proposed scheme. However, when $R_{req}>5 $ bits/s/Hz, the performance of the channel-priority-based scheme experiences severe degradation. Under the initial equal power allocation, the ergodic rate of the interfered user is $5.32$ bits/s/Hz. Therefore, when $R_{req}\leqslant 5 $ bits/s/Hz, the BS power is reduced. Under the channel-priority-based scheme, the BS causing stronger interference to radars has its power reduced more. In our topology, the BS causing stronger interference to radars has a poorer channel to the user. Reducing its power greatly lowers the interference to radars while causing small impacts on the user. In contrast, when $R_{req}>5$ bits/s/Hz, the radar power is reduced. Under the channel-priority-based scheme, the radar causing stronger interference to the user has its power reduced more. In our topology, the radar that has the best interference channel to the user is the radar that suffers from the highest communication interference. Although reducing its power greatly reduces the interference,  the detection performance of this radar is severely impacted, and consequently, impairs the overall detection performance. Therefore, the power allocation between C\&S systems cannot simply consider one factor in isolation. Our proposed scheme  provides a solution to consider two-sided impacts of the power in the spectrum-sharing context.

\begin{figure}[t]
	\centering
	\includegraphics[width=0.42\textwidth]{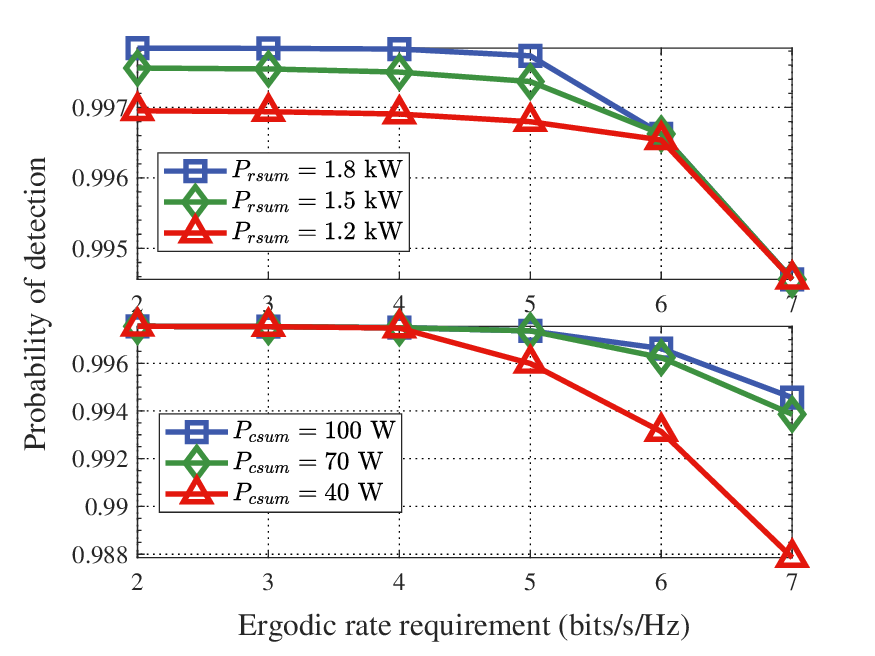}
	\caption{The radar detection performance under different power budgets, varies with the ergodic rate requirement.}
	\label{fig3}
\end{figure}

	As for the unilateral schemes of optimizing $\mathbf{P}_c$ or $\mathbf{P}_r$ only. The equal power allocation was used for the not optimized side, i.e., $\mathbf{P}_r=\min\{\frac{P_{rsum}}{M_r},P_{rmax}\}\mathbf{I}_{M_r}$  for  optimizing $\mathbf{P}_c$ and $\mathbf{P}_c=\min\{\frac{P_{csum}}{M_c}, P_{cmax}\}\mathbf{I}_{M_c}$ for  optimizing $\mathbf{P}_r$. The results of the optimized side are obtained by {\bf{Algorithm 1}}.  From Fig. \ref{fig2}, we can see that, the efficacy of unilateral schemes is constrained by limited DoFs. Only optimizing $\mathbf{P}_c$ fails to provide solutions when $R_{req}\geqslant 6$ bits/s/Hz. We provide a numerical explanation for this phenomenon. When the radar side does not control its transmit power, there is around a tenfold increase in the interference-plus-noise suffered by the user, i.e., from $-107$ dBm to $-98.9$ dBm.  Under this condition, the limit ergodic rate is $5.66$ bits/s/Hz, which is obtained by solving the ergodic-rate maximization problem \cite{reff}.	This indicates that when the radar side does not control its power, no matter how the power is allocated among BSs, the ergodic rate of the interfered user can not exceed $5.66$ bits/s/Hz. This explains why the curve of optimizing $\mathbf{P}_c$ only is missing in the region of $R_{req}\geqslant 6$ bits/s/Hz.  Therefore, the joint power allocation is more effective in providing scalable C\&S services.

\begin{figure}[t]
	\centering
	\includegraphics[width=3.0in]{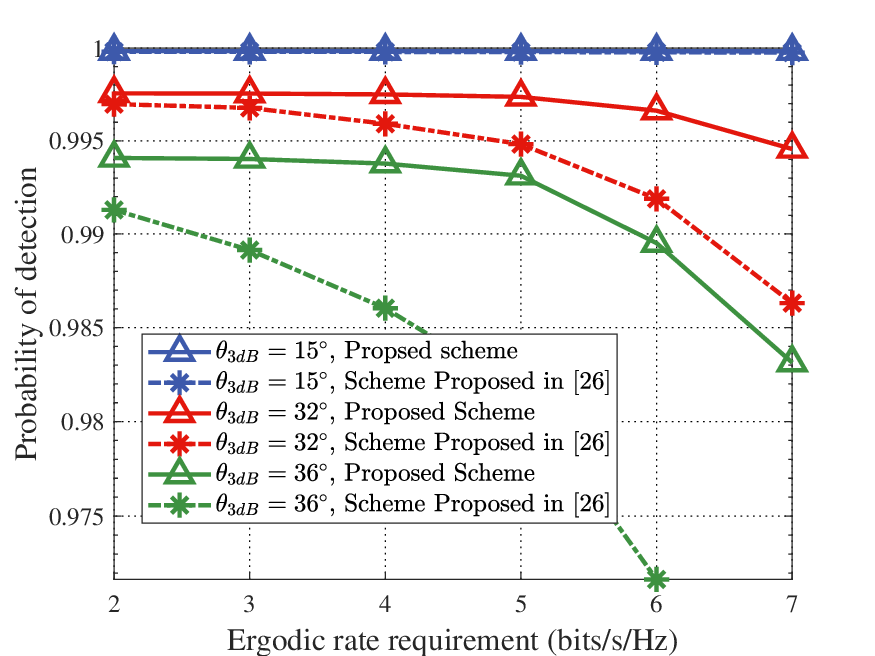}
	\caption{The radar detection performance under different half-power beam widths, varies with the ergodic rate requirement.}
	\label{fig4}
\end{figure}

	In Fig. \ref{fig3}, we show the impact of the power budgets, i.e.,  $P_{csum}$ and $P_{rsum}$, on C\&S performance.  It is evident that the detection performance improves with more power resources.  In addition, we  find that the curves under different $P_{csum}$ overlap when the ergodic rate requirement of the interfered user, i.e., $R_{req}$, is low, and the curves under different $P_{rsum}$ overlap when  $R_{req}$ is high. The reason is that, when $R_{req}$ is low, the communication side can meet the ergodic rate requirement without utilizing its full power. To maximize the radar SINR, the communication side reserves  its power and the sensing side operates at full capacity.
	Consequently, the curves under different $P_{csum}$ overlap.  In contrast, when $R_{req}$ is high, the sensing side has to control its power and the communication side works at full capacity. This leads to the  curves of different $P_{rsum}$ overlapping in the high $R_{req}$ region. We further find that, in spectrum sharing scenarios where two systems are mutually restricted, the optimal solution lies at the boundary of the feasible set but not within it. In other words, one system must exhaust its resources in the optimal solution.

	In Fig. \ref{fig4}, we show the impact of radar antenna directivity,  depicted by the half-power beam width, i.e., $\theta_{\text{3dB}}$, on the system performance. It can be seen that the detection probability experiences severe degradation as the half-power beam width increases. This is because the larger beam width leads to more power leakage to the interfered user and also opens a larger hole to receive BS interference.  In addition, the results show that the gap between the proposed scheme and the scheme proposed in \cite{ref11} is more obvious when $\theta_{\text{3dB}}$ is large.
	The reason is that, when $\theta_{\text{3dB}}$ is small, the C\&S signals are nearly decoupled by directional beams. The improved space left for the proposed scheme is restricted compared to \cite{ref11}.
	In contrast, when $\theta_{\text{3dB}}$ is large, C\&S signals are mutually coupled. The proposed scheme functions like directive beams to separate C\&S signals in the spatial domain. Consequently, it significantly outperforms the scheme proposed in \cite{ref11}. In fact, the proposed large-scale-CSI-based scheme works to optimize the spatial distribution of power to mitigate C\&S interference.

	 	\begin{figure}[t]
	 		\centering
	 		\includegraphics[width=0.48\textwidth]{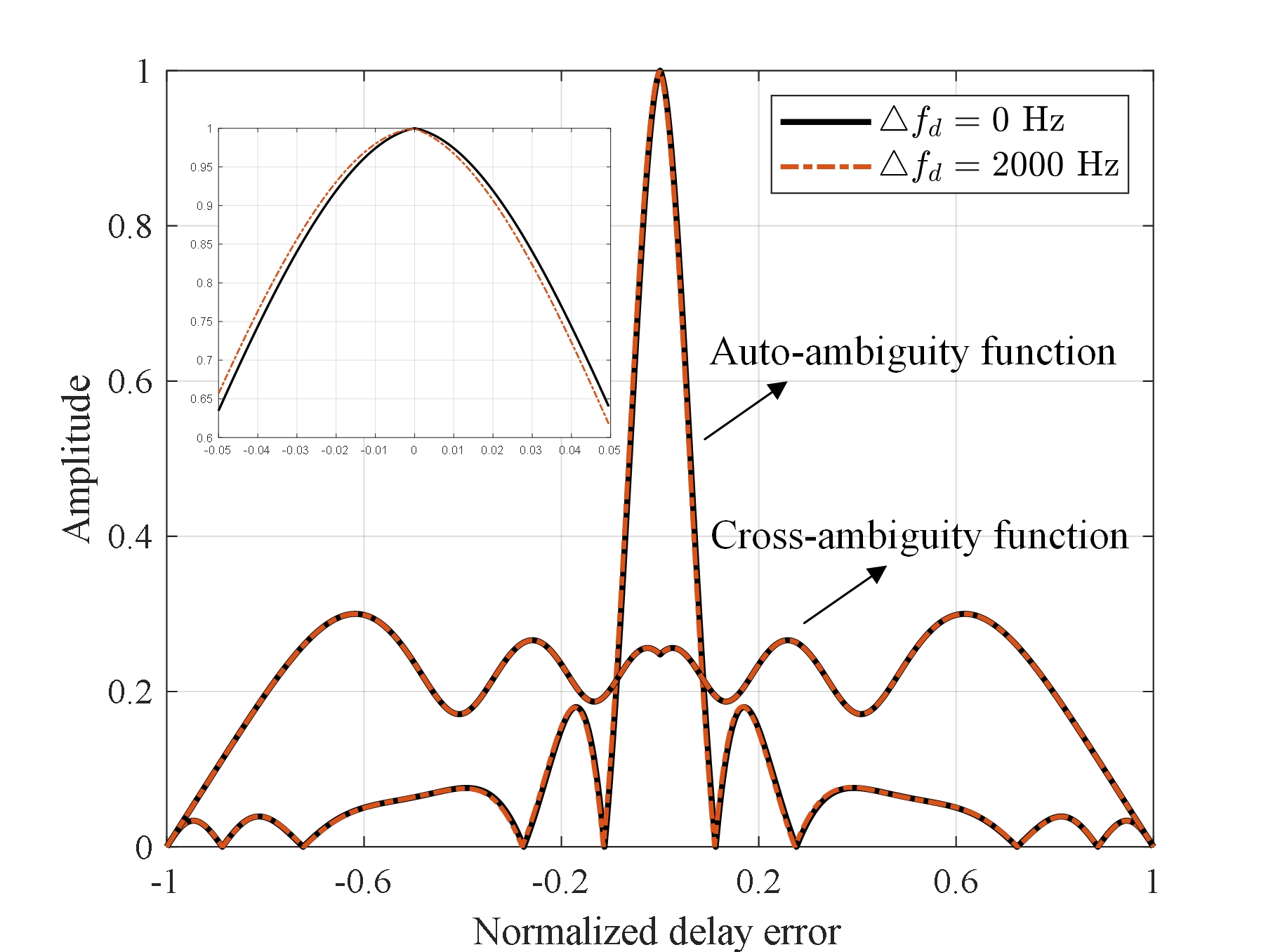}
	 		\caption{Zero-Doppler shift and max-Doppler shift cuts of the auto- and cross- ambiguity functions.}
	 		\label{fig5}
	 	\end{figure}

\begin{figure}[t]
	\centering
	\includegraphics[width=0.48\textwidth]{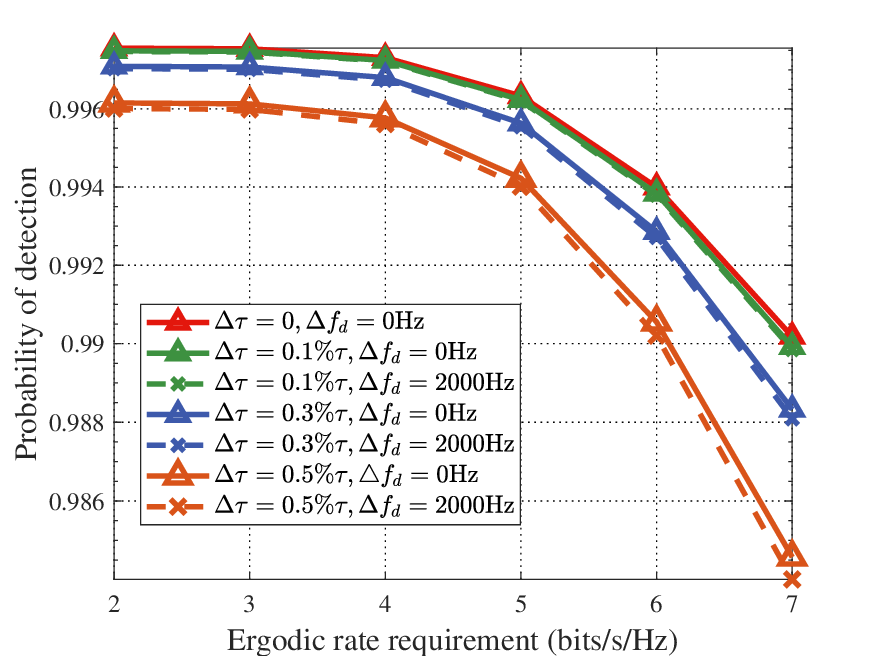}
	\caption{The radar detection performance under different delay and Doppler shift errors, varies with the ergodic rate requirement.}
	\label{fig7}
\end{figure}

	Finally, we discuss the impacts of the delay and Doppler shift errors on the radar detection. Specifically, we apply the single-band chirp waveforms as the radar waveforms: $u_1(t)=\frac{1}{\sqrt{\tau}}e^{j(\pi B\frac{t^2}{\tau}-\pi Bt)},  u_2(t)=\frac{1}{\sqrt{\tau}}e^{j(-\pi B\frac{t^2}{\tau}+\pi Bt)}, t\in[0,\tau]$ \cite{ad5}. The auto- and cross- ambiguity functions of the above two waveforms are given by
	 	 \begin{align} \nonumber
	 	 &\chi_{1,1}(\Delta{\tau},\Delta{f_d})=\\ \nonumber
	 	&\quad\quad\quad \left\{
	 	\begin{aligned}
	 		&\left|\frac{\sin[\pi g(\Delta{f_d})(\tau+\Delta{\tau})]}{\pi g(\Delta{f_d})\tau}\right|,  &\Delta\tau\in[-\tau, 0]\\
	 		&\left|\frac{\sin[\pi g(\Delta{f_d})(\tau-\Delta{\tau})]}{\pi g(\Delta{f_d})\tau}\right|,  &\Delta\tau\in[0, \tau]
	 	\end{aligned}
	 	\right.
	 \end{align}
	 where $g(\Delta{f_d})\triangleq \frac{B}{\tau}+\Delta{f_d}$ and $B$ is the waveform bandwidth.

	\begin{align}
		\nonumber
	 	& \chi_{1,2}(\Delta\tau,\Delta{f_d})=\\ \nonumber
	 	&\left\{
	 	\begin{aligned}
	 		&\left|g(\Delta \tau)\int_{0}^{\tau+\Delta \tau}e^{j2\pi(\Delta f_d+\frac{u-\Delta \tau-\tau}{\tau}B)u}du\right|, &\Delta \tau\in[-\tau, 0]\\
	 		&\left|g(\Delta \tau)\int_{\Delta \tau}^{\tau}e^{j2\pi(\Delta f_d+\frac{u-\Delta \tau-\tau}{\tau}B)u}du\right|, &\Delta \tau\in[0, \tau]
	 	\end{aligned}
	 	\right.
	 \end{align}
	where $g(\Delta \tau)\triangleq e^{j(\pi B\frac{\Delta \tau ^2}{\tau}+\pi B\Delta \tau)}. $
	 We set the pulse duration to $\tau=2$ us, the bandwidth to $B=5$ MHz, and the Doppler shift to $f_d=2000$ Hz, corresponding to an approximate velocity of $v \approx214$ m/s. In Fig. \ref{fig5}, we present the zero-Doppler shift and max-Doppler shift cuts of the auto- and cross- ambiguity functions. It can be seen that the curves representing the zero-Doppler shift  and max-Doppler shift, i.e., $\Delta f_d=0$ Hz and $\Delta f_d=2000$ Hz, are nearly overlapped. In contrast, the ambiguity functions change rapidly with delay errors. This indicates that the timing inaccuracy is the critical factor that impacts the matched filtering.
	To evaluate this impact, we present the detection performance of the proposed scheme under different delay and Doppler shift errors in Fig. \ref{fig7}. It can be seen that the curves representing different Doppler shift errors are closely overlapped while the curves associated with different delay errors demonstrate a noticeable discrepancy, with a 0.5\% delay error bringing 6\% detection accuracy degradation. Therefore, aligning delay discrepancies emerges as a pivotal consideration for the distributed radar sensing. Potential solutions include incorporating delay-insensitive detectors and advanced delay compensation algorithms within our framework, which deserve our further studies.

		\section{CONCLUSIONS}
		\label{sec5}
		In this paper, we have addressed C\&S interference within a distributed network. Given that the available C\&S cooperation is limited in the real world, we have proposed a radio-map-based C\&S cooperation framework. The radio map has been used to estimate the large-scale CSI so that the pilot-based C\&S interactions are bypassed. With the large-scale CSI, we have devised a joint power allocation scheme to maximize the radar detection performance while ensuring the ergodic rate requirement. To handle the non-convex C\&S metrics, we have employed the techniques of auxiliary-function-based scaling and fractional programming. The proposed problem has been effectively solved using an iterative algorithm. Simulation results have verified that the extrinsic information, i.e., positions and surroundings, is effective in coordinating conflicting signals. The proposed scheme provides a viable solution for the coexistence of C\&S systems in a loosely cooperated manner.

		\ifCLASSOPTIONcaptionsoff
		\newpage
		\fi
\appendices
\section{PROOF OF LEMMA}
	We calculate  the derivative of $g(\mathbf{P}_c, \mathbf{P}_r, z)$,
	\begin{equation}
	 \begin{aligned}
				\frac{\partial g(\mathbf{P}_c, \mathbf{P}_r, z)}{\partial z}&=\frac{\log_2(e)}{z}\bigg[N_c-\sum_{j=1}^{M_c}\big(\frac{N_cl_{c_j}p_{c_j}}{z\sigma^2_c(\mathbf{P}_r)+N_cl_{c_j}p_{c_j}}\big)^2\bigg]\\
				&> \frac{\log_2(e)}{z}(N_c-M_c). \nonumber
	\end{aligned}
    \end{equation}
	Therefore, under the condition of $N_c\geqslant M_c$, $g(\mathbf{P}_c, \mathbf{P}_r, z)$ is monotonically increasing with $z$. Since we have $g(\mathbf{P}_c, \mathbf{P}_r, z)=\bar{R}_{ap}(\mathbf{P}_c, \mathbf{P}_r,v^*)$ when $z=v^*$, it is obvious that the \textbf{Lemma} holds if $g(\mathbf{P}_c, \mathbf{P}_r, z)$ is monotonically increasing with $z$.

\section{PROOF OF CONVERGENCE OF THE ITERATIVE ALGORITHM}
For the convenience of expression, we introduce a function $f(\mathbf{P}_c,p_{r_i},\beta_i)$ to represent the left side of \eqref{41}:
\begin{equation}
	f(\mathbf{P}_c,p_{r_i},\beta_i)=2\beta_i\sqrt{[h_{{r_i}\rightarrow {r_i}}]^Hp_{r_i}h_{{r_i}\rightarrow {r_i}}}-(\beta_i)^2\sigma^2_{r_i}(\mathbf{P}_c).\nonumber
\end{equation}
Therefore, the objective of the proposed iterative algorithm can be rewritten as
\begin{equation}
	\max\limits_{\mathbf{P}_c, \mathbf{P}_{r}} \min\limits_i f(\mathbf{P}_c,p_{r_i},(\beta_i^*)^{s-1}).
\end{equation}
Define $\{(\mathbf{P}_c)^{s},(\mathbf{P}_r)^{s}\}$ as the optimal solution obtained in iteration $s$. Then, we have that
	\begin{equation}
	\label{eq114}
	\begin{aligned}
	\min\limits_if\big((&\mathbf{P}_c)^{s},(p_{r_i})^{s},(\beta_i^*)^{s-1}\big)\\
	&=\max\limits_{\mathbf{P}_c, \mathbf{P}_{r}} \ \big[\min\limits_if(\mathbf{P}_c,p_{r_i},(\beta_i^*)^{s-1})\big] \\
	&\geqslant \min\limits_if\big((\mathbf{P}_c)^{s-1},(p_{r_i})^{s-1},(\beta_i^*)^{s-1}\big).
	\end{aligned}
	\end{equation}
In addition, since $(\beta_i^*)^{s-1}$ is the optimal solution to $\max\limits_{\beta_i} f\big((\mathbf{P}_c)^{s-1},(p_{r_i})^{s-1},\beta_i\big)$, we can derive the following inequalities:
	\begin{equation}
	\label{eq115}
	\begin{aligned}
	\min\limits_if\big((\mathbf{P}_c)^{s-1},&(p_{r_i})^{s-1},(\beta_i^*)^{s-1}\big)\\
	&=\max\limits_{\beta_i}\big[\min\limits_i f\big((\mathbf{P}_c)^{s-1},(p_{r_i})^{s-1},\beta_i\big)\big]\\
	&\geqslant \min\limits_if\big((\mathbf{P}_c)^{s-1},(p_{r_i})^{s-1},(\beta_i^*)^{s-2}\big).
	\end{aligned}
	\end{equation}
Combining \eqref{eq114} and \eqref{eq115}, it is easy to conclude that the value of the objective is at least not decreasing in the iterative process:
	\begin{equation}
	\begin{aligned}
	\min\limits_if\big((\mathbf{P})^{s}_c,&(p_{r_i})^{s},(\beta_i^*)^{s-1}\big)\\
	&\quad\quad\geqslant\min\limits_if((\mathbf{P}_c)^{s-1},(p_{r_i})^{s-1},(\beta_i^*)^{s-2}).
	\end{aligned}
	\end{equation}
According to the monotonous boundary theorem, the proposed iterative algorithm is assured to be convergent.

\end{document}